\documentclass[pra,aps,twocolumn,nopacs,superscriptaddress,nofootinbib,longbibliography]{revtex4-1}
\usepackage{amsmath}  \usepackage{amssymb}  \usepackage{amsfonts}  \usepackage{bm}  \usepackage{bbm}   \usepackage{braket}  \usepackage{color}  \usepackage{comment}  \usepackage{dcolumn}  \usepackage{enumerate}  \usepackage{epsfig}  \usepackage{gensymb}  \usepackage{graphicx}  \usepackage{indentfirst}  \usepackage{lmodern}  \usepackage{mathrsfs}  \usepackage{mathtools}  \usepackage{psfrag}  \usepackage{pst-all}  \usepackage{soul}  \usepackage{units}  \usepackage{xcolor}
\usepackage{float}
\usepackage[colorlinks,linkcolor=blue,citecolor=blue,urlcolor=blue,hyperindex,driverfallback=dvipdfm]{hyperref}  \usepackage[T1]{fontenc}

\def\ii{{\rm i}}  \def\ee{{\rm e}}

      \def\Rb{{\bf R}}

\def\kb{{\bf k}}        
              
      \def\0b{{\bf 0}}

\def\mub{{\boldsymbol \mu}}

  \def\kB{{k_{\rm B}}}

\def\Eb{{\bf E}}

\newcommand{\energy}{\mathcal{E}}
    
\def\EF{{\energy_{\rm F}}}

\def\wp{\omega_{\rm p}}

\def\eps{\epsilon}    % definition of the permittivities
\def\epsm{\eps_{\rm m}}
\def\rdm{r_{\rm dm}}    \def\rmd{r_{\rm md}}
\def\tdm{t_{\rm dm}}    \def\tmd{t_{\rm md}}
\def\epsd{{\eps_{\rm d}}} \def\dm{d} % \def\dm{d_{\rm m}} 

\def\out{{\rm out}}    
\def\ww{\omega}  \def\inn{{\rm in}}
\def\ex{{\rm ex}}

\def\sp{{\rm sp}}

\def\Gg{\mathcal{G}}
             \def\log{\text{log}}

\newcommand{\imag}[1] {\mathopen{}{\rm Im}\left\{#1\right\}\mathclose{}}

\begin{document}
\title{Nonlinear photoluminescence in gold thin films}

\author{A.~Rodr\'{\i}guez~Echarri}
\affiliation{ICFO--Institut de Ciencies Fotoniques, The Barcelona Institute of Science and Technology, 08860 Castelldefels (Barcelona), Spain}

\author{F.~Iyikanat}
\affiliation{ICFO--Institut de Ciencies Fotoniques, The Barcelona Institute of Science and Technology, 08860 Castelldefels (Barcelona), Spain}

\author{S.~Boroviks}
\affiliation{Center for Nano Optics, University of Southern Denmark, Campusvej 55, DK-5230 Odense M, Denmark}
\affiliation{Nanophotonics and Metrology Laboratory, Swiss Federal Institute of Technology Lausanne (EPFL), Station 11, CH 1015, Lausanne, Switzerland}

\author{N.~Asger~Mortensen}
\affiliation{POLIMA--Center for Polariton-driven Light--Matter Interactions, University of Southern Denmark, Campusvej 55, DK-5230 Odense M, Denmark}
\affiliation{Center for Nano Optics, University of Southern Denmark, Campusvej 55, DK-5230 Odense M, Denmark}
\affiliation{Danish Institute for Advanced Study, University of Southern Denmark, Campusvej 55, DK-5230 Odense M, Denmark}

\author{Joel~D.~Cox}
\affiliation{POLIMA--Center for Polariton-driven Light--Matter Interactions, University of Southern Denmark, Campusvej 55, DK-5230 Odense M, Denmark}
\affiliation{Center for Nano Optics, University of Southern Denmark, Campusvej 55, DK-5230 Odense M, Denmark}
\affiliation{Danish Institute for Advanced Study, University of Southern Denmark, Campusvej 55, DK-5230 Odense M, Denmark}

\author{F.~Javier~Garc\'{\i}a~de~Abajo}
\email{javier.garciadeabajo@nanophotonics.es}
\affiliation{ICFO--Institut de Ciencies Fotoniques, The Barcelona Institute of Science and Technology, 08860 Castelldefels (Barcelona), Spain}
\affiliation{ICREA--Instituci\'o Catalana de Recerca i Estudis Avan\c{c}ats, Passeig Llu\'{\i}s Companys 23, 08010 Barcelona, Spain}

% --- abstract --------------------------------------------
\begin{abstract}
Promising applications in photonics are driven by the ability to fabricate crystal-quality metal thin films of controlled thickness down to a few nanometers. In particular, these materials exhibit a highly nonlinear response to optical fields owing to the induced ultrafast electron dynamics, which is however poorly understood on such mesoscopic length scales. Here, we reveal a new mechanism that controls the nonlinear optical response of thin metallic films, dominated by ultrafast electronic heat transport when the thickness is sufficiently small. By experimentally and theoretically studying electronic transport in such materials, we explain the observed temporal evolution of photoluminescence in pump-probe measurements that we report for crystalline gold flakes. Incorporating a first-principles description of the electronic band structures, we model electronic transport and find that ultrafast thermal dynamics plays a pivotal role in determining the strength and time-dependent characteristics of the nonlinear photoluminescence signal, which is largely influenced by the distribution of hot electrons and holes, subject to diffusion across the film as well as relaxation to lattice modes. Our findings introduce conceptually novel elements triggering the nonlinear optical response of nanoscale materials while suggesting additional ways to control and leverage hot carrier distributions in metallic films.
\end{abstract}
\maketitle
\date{\today}

% =========================================================
% --- introduction ----------------------------------------
% =========================================================
\section{Introduction}

Photoluminescence (PL)---the inelastic emission of light from metals produced upon recombination of photo-excited charge carriers---provides a probe of the electronic structure of materials and their non-equilibrium dynamics following optical pumping~\cite{M1969}. This phenomenon finds practical application in labelling of biological samples~\cite{FBC05}, mapping the near fields associated with plasmon resonances~\cite{BBN03_1,INO04}, and microscopy~\cite{YOT03,S20}. In this context, multiphoton or nonlinear PL (NPL), resulting in the emission of photons at energies exceeding (but, in general, not being a multiple of) that of the pumping ones, provides improved spatial resolution beyond the diffraction limit through the nonlinear dependence on the excitation field strength, while also probing a wider range of electronic states around the Fermi energy that reveal a rich interplay of energetic electrons and holes~\cite{PO97}.

Metallic thin films constitute an appealing playground to investigate ultrafast charge dynamics, with potential applications enabled by the modification of their electronic structure produced when the thickness goes down to a few atomic layers~\cite{paper335}. In this regime, quantum confinement effects can potentially tailor their optical and thermal properties~\cite{DC1987,QXL15,QXL16}. Photoluminescence constitutes a natural way of probing these effects~\cite{GFK19}, which become increasingly important as the metal thickness approaches the ultrathin regime.

% ---------------------------------------------------------
\begin{figure*}
    \includegraphics[width=1\textwidth]{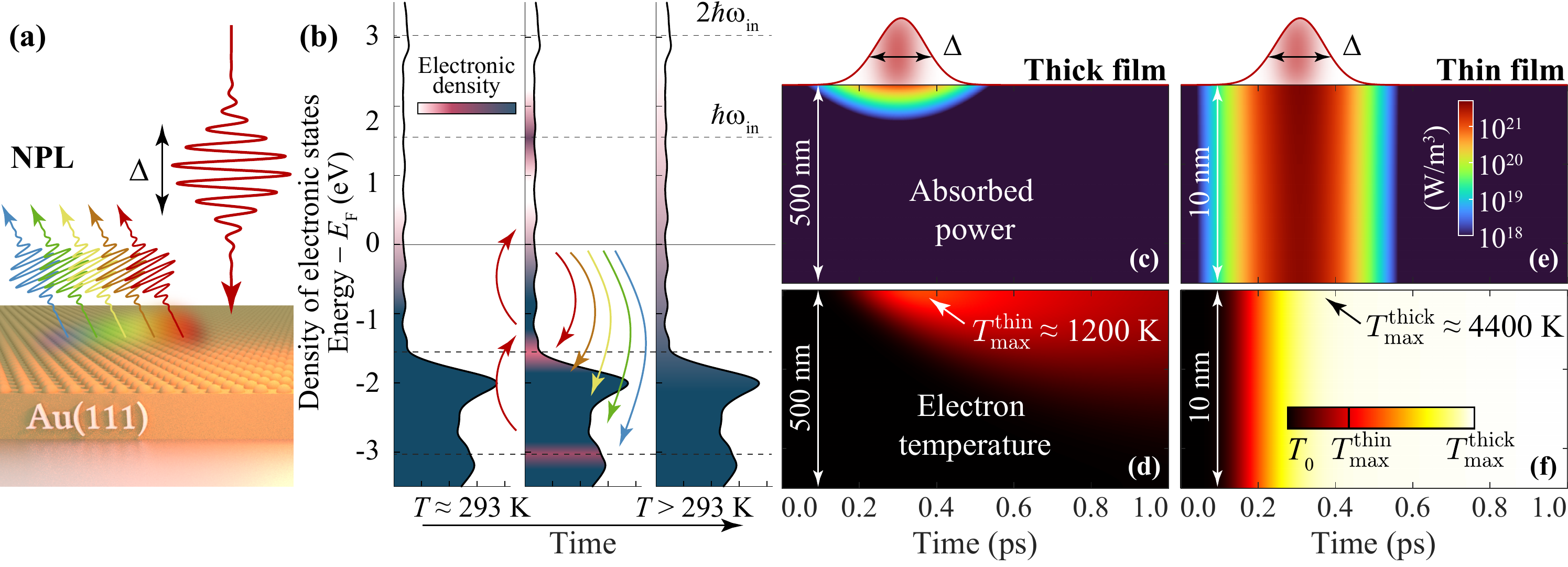}
    \caption{
    {\bf Optical excitation and electron thermalization in optically pumped gold films.}
    (a) Illustration of a Au(111) thin film illuminated by a femtosecond Gaussian pulse, along with the resulting TPP emission that we study in this work. 
    (b) Dynamics of the occupied density of electronic states upon optical excitation. From the equilibrium state at room temperature, electrons are promoted to higher levels, and the associated holes are subsequently filled by emitting photons with different energies (colored curves). Once the pulse is gone, after a transient time of $\sim$ 10's femtoseconds, the electronic states reach thermal equilibrium at a higher temperature than initially.
    (c,d) Time and depth dependence of (c) the absorbed power and (d) the electronic temperature in a 500\,nm film. (e,f) Same as (c,d), but for a 10\,nm film.}
    \label{Fig1}
\end{figure*}

The excitation of a metal surface with intense optical fields gives rise to nonlinear optical phenomena such as second-harmonic generation (SHG) and two-photon absorption~\cite{CCS1981,HBA07,SLZ12}, the former being a coherent process, while the latter exhibits an incoherent behavior that manifests in the two-photon photoluminescence (TPP) emission~\cite{JPJ13,XLG19}. More specifically, SHG is a second-order instantaneous nonlinear optical process that is spectrally concentrated around the second-harmonic frequency of the excitation field, whereas TPP is associated with a broad, weaker spectral signature imbued with features of the electronic band structure~\cite{BYS1986,INO04,INO05,BBL05}. While SHG relies on the breaking of inversion symmetry at the surface of centrosymmetric noble metals, two-photon absorption is a third-order process that can occur in the bulk of the metal~\cite{SH00} and involves contributions from a wide range of electronic states and transition energies. Both SHG and TPP are sensitive to the crystallinity of the surface~\cite{CCS1981,BCS09,paper363}. The nonlinear dependence on the electric-field amplitude $E$ of the incident light inside the metal thus leads to an initial onset of $E^4$ TPP scaling (quadratic in the input intensity), with larger incident intensities triggering higher-order multiphoton processes that scale with higher powers of $E$ that contribute to overall NPL~\cite{MEM05,BBS12}.

The mechanism responsible for TPP in gold was first attributed to radiative recombination of holes in the valence d-band with electrons in the conduction band~\cite{CCS1981,BCS09,BBH12,BMA15}. Subsequent experiments consolidated a three-step model of radiative recombination following two linear absorption events that promote d-band electrons to the conduction band, although some debate as to the nature of TPP in the infrared range of the electromagnetic spectrum persists~\cite{BBN03_1}. The important role of the electronic structure in the NPL process has stimulated explorations of the mechanism in mesoscopic gold films of 10--100\,nm thickness~\cite{GFK19}, where confinement effects in the electronic spectrum lead to a NPL intensity scaling beyond the $E^4$ rule. In this respect, it is important to note that thermal effects in non-equilibrium systems~\cite{paper280,DS19}, which are associated with elevated electronic temperatures following thermalization on ultrafast (tens of femtoseconds) timescales, play an important role in the PL signatures of metallic structures, as has been demonstrated for gold nanoparticles~\cite{HKB15}. In particular, the connection between photothermal effects and the light emitted from metal films can be well-understood in the context of the Fowler--DuBridge model~\cite{F1931,D1932,D1933,ZZ20}, which predicts a quadratic scaling of the PL signal with temperature that is found to increase for thin films as compared to semi-infinite samples under the same illumination conditions. However, these previous studies ignore electronic transport across nanometer distances, which we find here to play a leading role in the strength and time-dependence characteristics of NPL.

We perform experiments on crystalline gold flakes and reveal the importance of electronic transport by developping a theoretical formalism to describe the interplay between electronic band structure and thermal effects in mesoscopic gold films when illuminated with a femtosecond pulse (see illustration in Fig.~\ref{Fig1}a). More precisely, we employ first-principles methods to simulate the electronic band structure of crystalline gold films, from which we calculate electronic transition rates and the electron contribution to the thermal heat capacity. We combine these elements with a two-temperature model (TTM) to incorporate thermal effects in the redistribution of electrons within bands around the Fermi level following irradiation with an intense optical pump pulse. The resulting spectrum of NPL emitted from thin films with increased surface-to-volume ratio is predicted to exhibit a much higher sensitivity to thermal effects, to which we attribute the breaking of the $E^4$ rule, while NPL from thick films is primarily controlled by the bulk electronic structure combined with the diffusion of thermal energy away from the surface. Our findings complement experimental efforts to understand light emission from gold nanoparticles~\cite{BCS09,MVP16,RKG17,JCS20}, while elucidating the role of the electronic temperature in recent NPL measurements of crystalline gold films~\cite{GFK19}. The present study provides a solid framework in combination with experiments to understand the TPP and thermo-optical properties of ultrathin metallic thin films, which constitute a key platform for diverse applications in nanophotonics.

% =========================================================
% --- results ---------------------------------------------
% =========================================================
\section{RESULTS AND DISCUSSION}  \label{sec:Results}

% ---------------------------------------------------------
\begin{figure*}
    \includegraphics[width=0.6\textwidth]{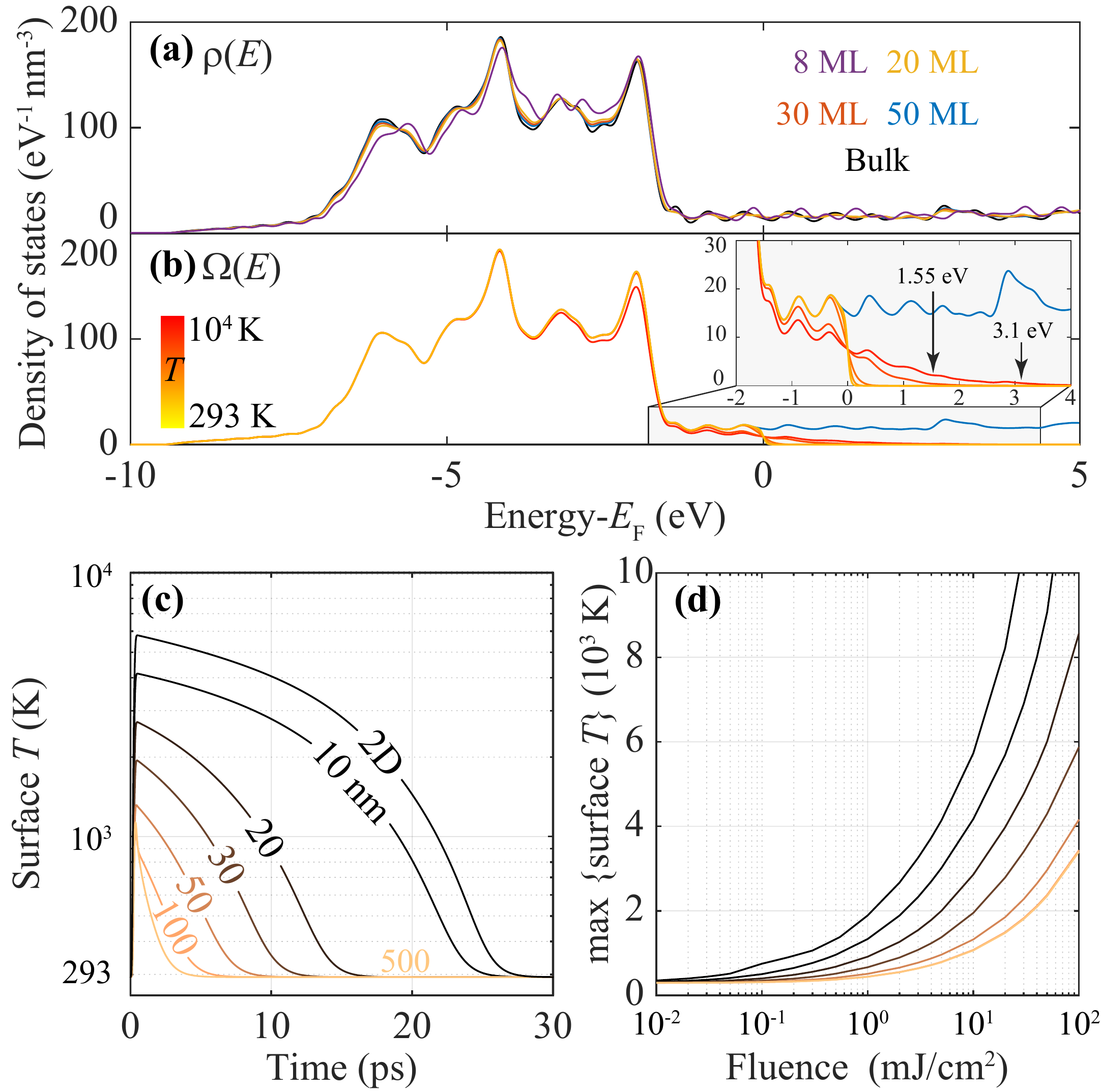}
    \caption{{\bf Temperature dynamics in optically pumped thin gold films}. (a) Density of states for few-atomic-thick films (see legend). 
    (b) Density of all states (blue curve) and density of occupied states as a function of temperature (colored curves) for a $N=50$~MLs gold film. The inset zooms into the region near the Fermi energy, where the arrows indicate the energy position of the impinging photons (i.e., $\hbar \ww_\inn = 1.55$\,eV) as well as its second harmonic. (c) Evolution of the electronic temperature at the surface for various film thicknesses under excitation with a pulse of 150\,fs and 10\,mJ/cm$^{2}$ fluence. 
    (d) Maximum surface electronic temperature as a function of fluence for the same pulse duration and different film thickness (same color code as in (c)).}
    \label{Fig2}
\end{figure*}

The current interest in the optical properties of ultrathin metal films~\cite{paper235,paper382,BS19} is partly due to the expectation that their modified electronic band structures give rise to different optical properties compared to their bulk counterparts. Finite-size and surface effects related to the crystalline quality of the films play a decisive role in the optical response~\cite{LTA14,QXL15}. As a motivation for the present study, we argue that NPL provides the means to interrogate the electron dynamics in such systems~\cite{PO97,KOP05}. Specifically, we are interested in modeling the NPL emission from free-standing gold films after being irradiated by femtosecond light pulses (see Fig.~\ref{Fig1}a), for which the emission profile is an integrated result depending on the available transitions within the gold band structure. We place special emphasis on understanding the NPL signal when varying the thickness of the gold films.  

Upon optical excitation, electrons in the metal partially absorb the pump energy, thus being heated to high temperatures, which together with direct multiphoton optical transitions are responsible for the resulting PL, as observed in gold nanoparticles~\cite{BCS09,RKG17,JCS20} (see Fig.~\ref{Fig1}b). In gold films, we assume that the pump beam is laterally wide enough to consider the problem to be one-dimensional, such that we only need to analyze the dependence on the $z$ coordinate along the out-of-plane direction. We simulate the depth and time dependence of the electron temperature $T(z,t)$ produced by a 150\,fs Gaussian pump pulse having a moderate fluence of $10$\,mJ/cm$^{2}$ and a central wavelength $\lambda_\inn=800$\,nm (i.e., $\hbar \ww_\inn \approx 1.55$\,eV) with the TTM given by~\cite{paper330,CXJ10,ZCH15,RIG17}
\begin{align} \label{Eq:TTM}
    C_e(T) \, \partial_t T = \partial_z\left[ k_e(T) \partial_z T \right] - G(T-T_l)+S,
\end{align}
where $C_e$ is the heat capacity, $k_e$ is the electron thermal conductivity, $G$ is the electron-phonon coupling coefficient, $T_l$ is the temperature of the lattice, and $S$ is the absorbed power from the optical femtosecond pulse (see the details in the Appendix). We find  the absorbed power to exhibit different characteristics for thick and thin films, as illustrated in Figs.~\ref{Fig1}c and \ref{Fig1}e for 500\,nm and 10\,nm thickness, respectively. In the thicker film, the optical pump field only penetrates a few tens of nanometers (as determined by the skin depth, or equivalently, the light propagation length $1/2k_\inn \imag{\sqrt{\epsm}}\sim 10$\,nm with $k_\inn = 2\pi /\lambda_\inn$), while a peak electron temperature $T_{\rm max}^{\rm thick}  \approx 1200$\,K is found at the top surface, partly attenuated by heat diffusion towards the bulk of the metal. In contrast, the same excitation in the thinner film, the excitation nearly uniformly permeates the entire metal volume because the thickness is no longer large compared with the skin depth, and optical surface reflections produce a more uniform field intensity. In addition, such reflections enhance the optical field in the metal, which together with the marginal role played by thermal diffusion, leads to a higher peak surface temperature $T_{\rm max}^{\rm thin} \approx 4400$\,K. The temperatures found in thin and thick films are both high enough to significantly affect the PL response because $\kB T_{\rm max}^{\rm thick} \approx 0.4$\,eV, so that electron-hole radiative recombination can occur with some probability by emitting photons of energy exceeding the pumping photon energy. This additional mechanism contributing to the PL at energies beyond the pump photons has an inelastic nature, in contrast to the few-transition processes associated with multiphoton absorption followed by radiative recombination (e.g., in TPP). As we shall see, these two mechanisms compete in importance depending on the thickness of the metal and the fluence of the pump.

Interestingly, finite-size effects associated with the thickness start to be observed below $\sim 10$\,nm, corresponding to $\sim 50$ gold atomic monolayers (MLs) with a (111) crystal orientation (considering an interlayer spacing of $\approx 0.24$\,nm). To illustrate this effect, Fig.~\ref{Fig2}a shows the 2D electronic density of states (DOS) $\rho(E)$ obtained for 8, 20, 30, and 50~MLs as well as for a semi-infinite surface. These results are computed from the band structure energies $\energy_{i,k}$, summed over the band index $i$ and electron wave vector $\kb$ (see eq~\eqref{Eq:DOS} in the Appendix). When approaching 50~MLs ($\sim 10$\,nm), the features in the DOS resemble those of the bulk, supporting the idea that band-structure phenomena become important only for few-atomic-thick films. Therefore, from the band structure viewpoint, gold films with thicknesses $\gtrsim 10$\,nm are considered to behave approximately equivalent to the bulk. However, for the thermal behavior of such films, as we have anticipated in Fig.~\ref{Fig1}, there exist large differences between 10\,nm and 500\,nm thick films regarding the temporal evolution and fluence dependence of the electron temperature. Importantly, note that we connect the TTM with the band structure of the film by computing the temperature-dependent chemical potential $\mu(T)$ from the DOS, and also the heat capacity that enters in the TTM analysis, see the Appendix and Fig.~\ref{FigS1} in Supplementary Figures.

The electronic bands are populated according to the Fermi--Dirac (FD) statistics, given by the electron energy $\energy$ and temperature $T$ as
\begin{align} \label{Eq:FD}
    f(\energy,T) = \left\{ \ee^{[\energy-\mu(T)]/\kB T}+1\right\}^{-1}.
\end{align}
We also define the occupation density $\Omega(\energy) = \rho(\energy) f(\energy,T)$, which varies accordingly with temperature (see Fig.~\ref{Fig2}b), such that a larger number of bands become populated at higher temperatures. Due to heating produced by optical absorption from the pump, the PL signal is dominated by electron dynamics near the surface, which is precisely where the temperature reaches its larger values. As shown in Fig.~\ref{Fig2}c, the surface electron temperature increases rapidly once the pulse hits the surface (see Fig.~S2 in SI). We observe two main effects: (1) thinner films reach higher temperatures compared with semi-infinite films; and (2) thinner films retain the temperature much longer than their bulkier counterparts. Thinner films boost the electric field intensities due to multiple reflections of the pump field inside the film (similar to a Fabry--P\'erot resonator), causing them to absorb a larger fraction of the pump energy and thus reach higher temperatures. Additionally, the fact that they are thinner reduces the influence of thermal diffusion in the TTM, hence removing such a cooling mechanism, which is in contrast significant for thicker films. 
The diminished role of the diffusion term implies that for thinner films the influence of the source is effectively instantaneous, and in the extreme case of a few atomic layers, diffusion (driven by the temperature gradient) can be ignored, giving rise to what we name 2D limit. Therefore, the range of surface temperatures varies from the 2D limit to the bulk limit, as illustrated in Fig.~\ref{Fig2}d. Films with an intermediate thickness (e.g., between 50\,nm and 200\,nm) present a hybrid behavior in which the temperature requires hundreds of nanoseconds to reach thermal equilibrium across the film thickness (see Fig.~\ref{Fig2} in the SI). 

The PL emission intensity is proportional to the number of photons that undergo transitions from excited states to lower-energy states by emitting a photon with energy $\hbar\omega_\out$. The spectral profile of the PL is thus given by
\begin{align} \label{Eq:I_TPA_out}
    I(\hbar\omega_\out) \propto \int_0^{t_{max}} dt\, \sum_\kb n_k(\hbar\omega_\out,\kb),
\end{align}
where the sum extends over the contributions of nearly vertical transitions (i.e., the photon energy is negligible compared to the size of the Brillouin zone) for each $\kb$-point in the electronic band structure and 
\begin{align} \label{Eq:nout_k}
    n_k(\hbar\omega_\out,\kb) = \sum_{ij} \gamma_{ij}^\sp (1-p_{i}) p_{j} \Gamma(\energy_{ji}-\hbar\ww_\out) \Theta_{ji}
\end{align}
is defined in terms of the radiative decay rate $\gamma_{ij}^\sp(\kb)$ (from band $j$ to band $i$ at a given $\kb$ point), the occupation factor $p_i(\kb)$ of the electronic state with energy $\energy_{i,\kb}$, the linewidth of the final state $\Gamma(\omega)$ (see eq~\eqref{Eq:smearing}), and the Heaviside distribution $\Theta_{ji} \equiv \Theta(\energy_{ji})$ with argument $\energy_{ji}=\energy_{j,\kb}-\energy_{i,\kb}$.
The electronic population $p_i(\kb)$ evolves with time according to the rate equation~\cite{paper280}
\begin{align} \label{Eq:rate_align}
    \frac{d p_{i}}{dt} =
    -\gamma^T &\left( p_{i}-p_{i}^{T} \right) +\sum_j \gamma^\ex_{ij} (p_{j}-p_{i})\\
    -\sum_{j=1}^{i-1} & \gamma^{\sp}_{ji}(1-p_{j}) p_{i} +\sum_{j=i+1} \gamma^{\sp}_{ij} (1-p_{i}) p_{j} , \nonumber
\end{align}
where $\gamma^T$ is a phenomenological damping rate, $p_i^T(\kb) = f(\energy_{i,\kb},T)$ is the thermalized population, and $\gamma_{ij}^{\rm ex}=\gamma^{(1)}_{ij}+\gamma^{(2)}_{ij}$ is the excitation composed of the linear absorption $\gamma^{(1)}(t) \propto E_0^2$ and $\gamma^{(2)}(t) \propto E_0^4$ to the fourth power, establishing the $E^4$ rule characteristic of the TPP (see the details in the Appendix). Note that all quantities depend on temperature, which evolves in time following the TTM given in eq~\eqref{Eq:TTM}.

% ---------------------------------------------------------
\begin{figure*}
    \includegraphics[width=1\textwidth]{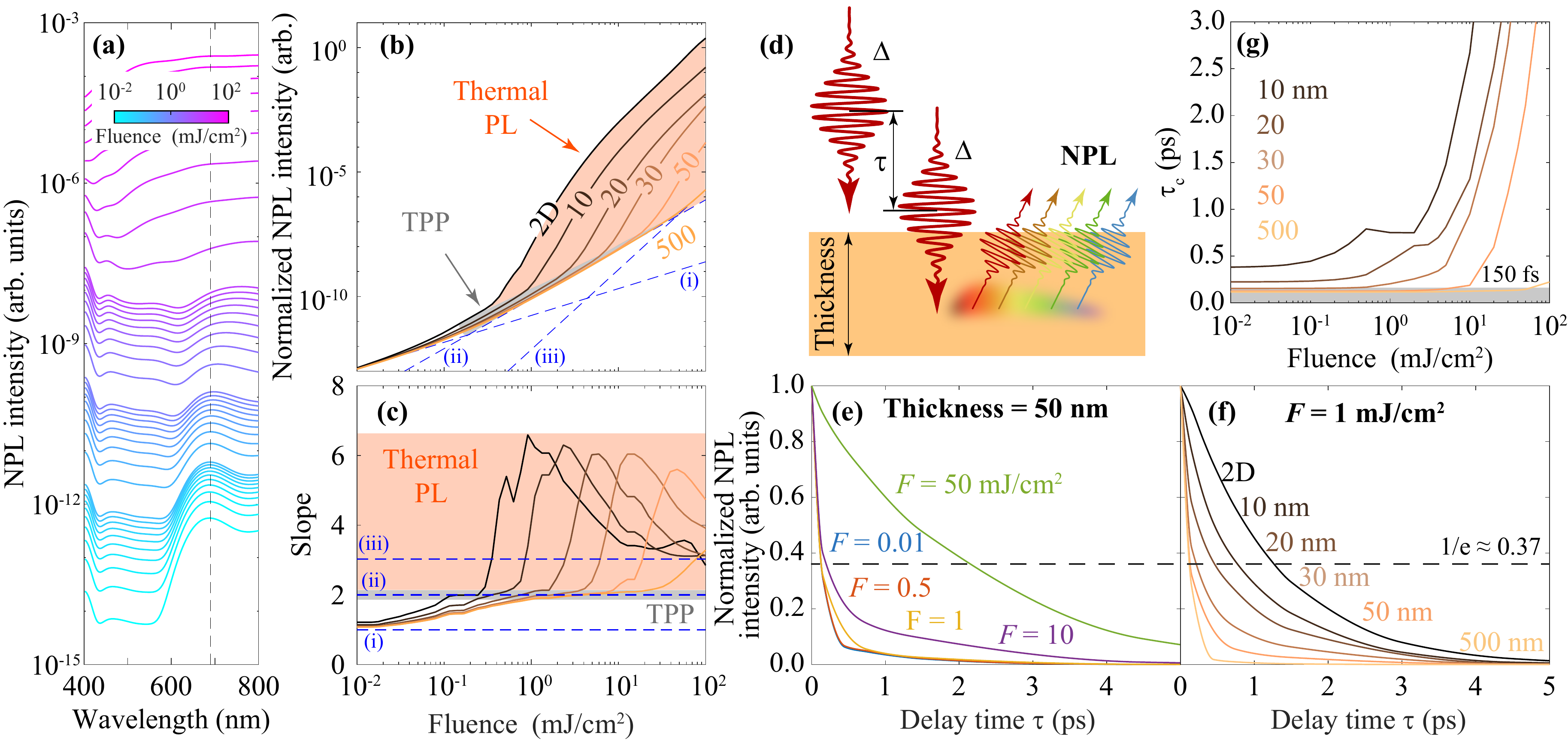}
    \caption{{\bf Nonlinear-photoluminescence (NPL) dependence on film thickness, pulse fluence, and two-pulse correlation.} (a) Spectral dependence of the NPL emission intensity for a film of 50\,nm thickness and a range of fluences indicated in the color scale. The vertical dashed line signals the emission peak at $\sim690$\,nm wavelength. 
    (b) Fluence dependence of the normalized NPL emission intensity for a range of thicknesses following the color-coordinated labels. The blue dashed lines serve as a guide for the linear (i), quadratic (ii), and cubic (iii) dependence of the NPL on fluence. In the narrow gray-shaded area, NPL exhibits a quadratic dependence on fluence, while in the orange area it is dominated by thermal effects (thermal PL). 
    (c) Derivative of the curves in (b) with respect to the fluence.
    (d) We perform two-pulse PL correlation simulations, assuming two identical pulses delayed by a time $\tau$ and impinging normally on a gold film of finite thickness. 
    (e,f) We study the two-pulse NPL correlation traces (normalized intensity) for (e) a film of 50\,nm thickness at several values of the fluence (see labels) and (f) a fixed fluence of 1\,mJ/cm$^2$ with several film thicknesses. The dashed line indicates a $1/\ee$ decay of intensity.
    (g) Characteristic $1/\ee$ decay time of the two-pulse correlation signal as a function of fluence for different film thicknesses. The shaded area indicates times below $150$\,fs.}
    \label{Fig3}
\end{figure*}

% ---------------------------------------------------------
\subsection{Thickness dependence of nonlinear photoluminescence}

In Fig.~\ref{Fig3}a, we plot the NPL spectral emission profile for a gold film of 50\,nm thickness calculated from eq~\eqref{Eq:I_TPA_out} for a set of fluences indicated in the legend (within a range spanning four orders of magnitude) using a pulse width $\Delta = 150$\,fs. We observe that, in this particular case, the resulting PL spectrum is peaked at $\sim 690$\,nm. When increasing the temperature, the emission peaks fade away to yield a broader spectrum.

To analyze the dependence of the NPL signal on thickness, we define the normalized intensity as
\begin{align}
    \bar{I} = \frac{1}{\ww_\inn} \int_{\ww_\inn}^{2\ww_\inn} d\ww_\out \, I(\hbar\omega_\out) ,
\end{align}
where the integration area covers the spectral region between the second-harmonic frequency $2\ww_\inn$ and the fundamental frequency $\ww_\inn=2\pi c/\lambda_\inn$. We compute the normalized intensity as a function of fluence in Fig.~\ref{Fig3}b for various values of film thickness, which clearly shows how the PL signal produced by thinner films reaches much higher values than that of thicker ones. For fluences $\lesssim 0.1$\,mJ/cm$^2$, the PL signal displays a linear dependence on pulse fluence, indicating that nonlinear absorption processes are not large enough to dominate the electronic response. However, for fluences between $\sim 0.1$\,mJ/cm$^2$ and $\sim 100$\,mJ/cm$^2$, the dependence for semi-infinite films is dominated by two-photon absorption following the $E^4$ rule and, therefore, being responsible for the TPP (i.e., a $F^2$ scaling with fluence). Incidentally, for any fluence within the $0.1-100$\,mJ/cm$^2$ range, the PL emission grows with decreasing film thickness faster than the square of the fluence, suggesting that thermal effects dominate. The gray shaded area highlights the region (denoted TPP) in which two-photon absorption dominates, whilst the orange area corresponds to dominance of thermal effects (denoted thermal PL). These regions are limited by the bulk (effectively investigated for a 500\,nm thick film) and 2D-thin film limits. We observe a gradual change in the PL intensity when the film thickness varies from 50\,nm to 10\,nm. The actual transition through the different regimes (NPL and thermal PL) might vary with the excitation wavelength $\lambda_\inn$, which is directly associated with the optical penetration depth $\sim 4\pi\imag{\epsm(\lambda_\inn)}/\lambda_\inn$ in which optical absorption takes place.

We identify the region in which the $E^4$ rule governs the emission by taking the derivative of the normalized intensity with respect to the incident fluence to compute its slope (see Fig.~\ref{Fig3}c). As a guide, we show blue dashed lines with slopes of 1, 2, and 3, respectively labeled (i), (ii), and (iii) in the figure. Within the grey shaded area, TPP dominates the NPL emission, which is overtaken by thermal PL when the electron temperature is elevated above a few thousands of degrees. We observe the transition from the TPP to thermal PL in Fig.~\ref{FigS3}a in Supplementary Figures, representing a fluence-thickness phase diagram, in which the same blue dashed curves corresponding to slopes of 1--3 are used to limit different regions. Additionally, Fig.~\ref{FigS3}b in Supplementary Figures shows a phase diagram of the maximum surface electron temperature, corroborating that thermal effects become relevant at temperatures larger than $T \gtrsim 10^3$\,K.

% ---------------------------------------------------------
\begin{figure*}
    \includegraphics[width=1\textwidth]{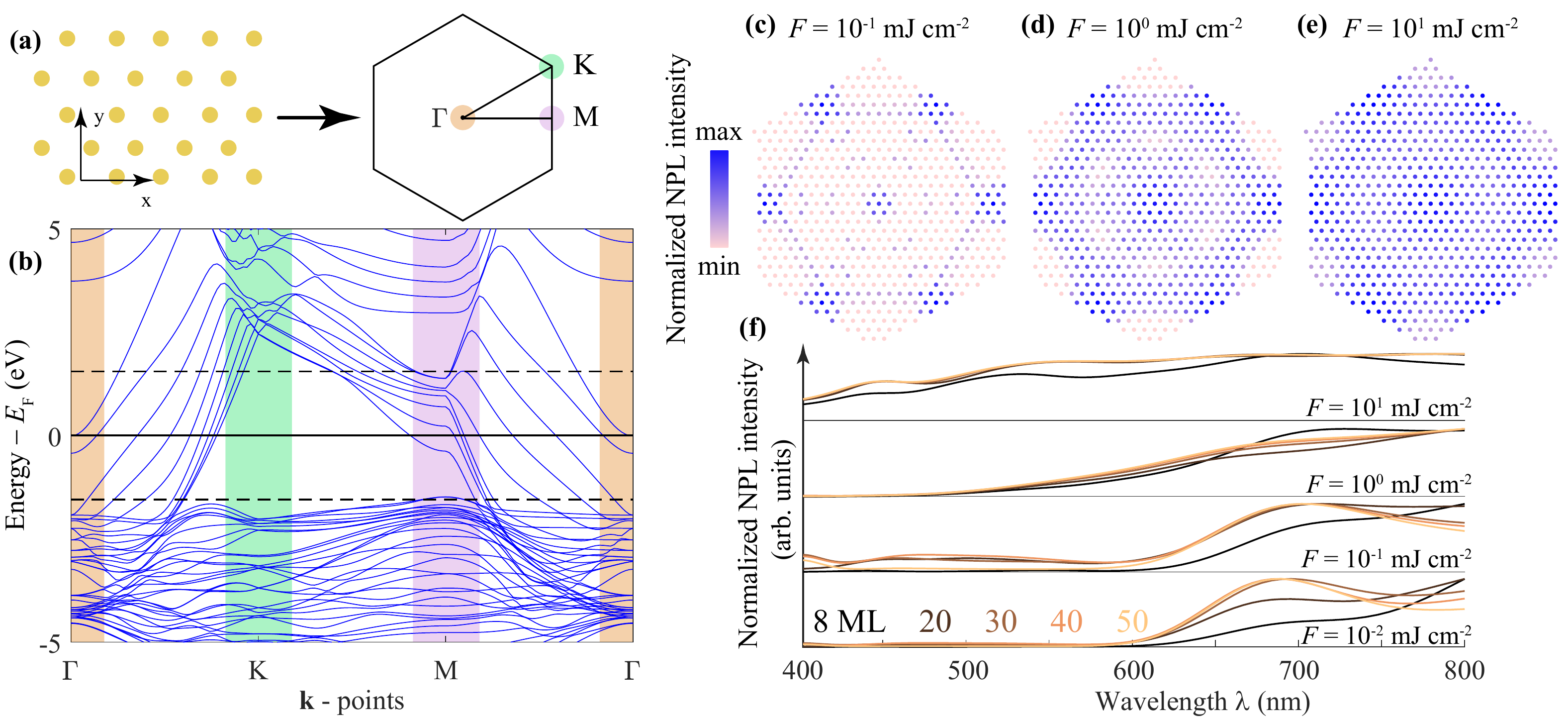}
    \caption{{\bf Au(111) band structure effects in NPL.} (a) Real space arrangement of the Au(111) atoms at the surface and the associated Brillouin zone, in which the highly symmetric $\Gamma$, K, and M points are highlighted by orange, green, and pink areas, respectively. (b) Electronic band dispersion of 8 ML Au(111) determined from first-principles calculations, where the three high-symmetry points are highlighted in colors according to the right sketch in panel (a). Panels (c-e) indicate the normalized NPL intensity corresponding to each of the points of the Brillouin zone for different values of the fluence as indicated in each panel. (f) Normalized NPL intensity for a discrete number of MLs, from 8 to 50 MLs, as indicated by the colored code, and for various values of the impinging fluence separated vertically.}
    \label{Fig4}
\end{figure*}

% ---------------------------------------------------------
\subsection{Two-pulse correlation in nonlinear photoluminescence}

As concluded above, the resulting NPL emission in gold films is driven by both TPP and thermal PL, depending on the pump fluence and film thickness. Each of these two mechanisms has its characteristic time scale, which can be interrogated by two-pulse excitation: we can study the NPL emission intensity when a copy of the pulse impinges on the sample at a time $\tau$ after the original one (see the sketch in Fig.~\ref{Fig3}d). Two-pulse correlation analysis is a powerful tool for studying the temporal electron dynamics, introducing a new parameter (i.e., the delay between the two copies of the pulse) besides the pulse intensity. The two pulses interact optically within a time window limited to the pulse width (i.e., $\tau_p = \Delta  = 150$\,fs in our study), whereas the characteristic time for thermal effects is given by $\tau_{\rm th} = C_e(T)/G$ according to eq~\eqref{Eq:TTM}. The latter spans a range from $\tau_{\rm th} \approx 1 $\,ps to $\tau_{\rm th} \approx 10 $\,ps for electron temperatures from $T=0.3\times 10^3$\,K to $T=3\times 10^3$\,K (see values of $\tau_{\rm th}(T)$ in Fig.~\ref{FigS4} in Supplementary Figures). 

The normalized intensity changes as shown in Fig.~\ref{Fig3}e for different values of the fluence in a film of 50\,nm thickness. For such a film thickness, fluences below 10\,mJ/cm$^2$ drive TPP, corresponding to values of 1/e decay below the picosecond (see Fig.~\ref{Fig3}g). However, for larger fluences (see green curve in Fig.~\ref{Fig3}e), the decay rate of the two-pulse correlation curve increases considerably to several picoseconds. Conversely, thermal effects dominate the PL signal at low fluences in thinner films, as shown in Fig.~\ref{Fig3}f for $F=1$\,mJ/cm$^2$ and several values of the film thickness. Corroborating this assignment, when thermal effects dominate the response, the decay rate is of the order of picoseconds and increases with increasing surface electronic temperature (i.e., for thinner films). In contrast, in the NPL regime, the decay rate is of the order of hundreds of femtoseconds and does not show significant dependence on fluence or temperature.

These results are in agreement with those discussed above for single pulses, in connection to Fig.~\ref{Fig3}b-c, and consistent with the fact that thinner films reach higher surface electron temperatures under the same fluence, and therefore, the NPL emission for thinner films extends over a longer time scale of the order of picoseconds. As a summary, we plot in Fig.~\ref{Fig3}g the characteristic decay time $\tau_{\rm c}$ as a function of fluence for various film thicknesses, showing that the width is of the order of hundreds of femtoseconds at fluences below a threshold value, characteristic of the given film thickness.

% ---------------------------------------------------------
\subsection{Band structure effects in nonlinear photoluminescence}

The emitted NPL corresponds to the collective emission of electrons from excited to lower energy states (see eq~\eqref{Eq:nout_k}). The electron energy levels are given in terms of in-plane crystal momentum from the computed band structure, which is sensitive to the crystallographic arrangement of atoms in the film. Figure~\ref{Fig4}a depicts the arrangement of atoms in a (111) crystallographic orientation and the associated Brillouin zone in 2D reciprocal space, where we identify the high-symmetry points $\Gamma$, M, and K. The band structure traversing these high-symmetry points is obtained from ab-initio calculations of a 10 ML Au(111) film and presented in Fig.~\ref{Fig4}b for energies relative to the Fermi energy $\EF$. Interestingly, the PL from each of these points in the band structure directly depends on the strength of the dipole transition matrix element $\mub_{ij}$ (see Appendix) between the involved states. In Fig.~\ref{Fig4}c-e we present the intensity of the emitted NPL for several fluences at the discretized points of the Brillouin zone to emphasize that there are preferable areas of emission in the reciprocal space. For low values of the fluence (panel (c)), and therefore, in the TPP regime, most of the emission correlates with the $\Gamma$ and M points. Additionally, there is a contribution from the $\Gamma-$K connection that correlates with the region in the band structure diagram in Fig.~\ref{Fig4}b, where the electronic bands cross the Fermi energy, in concordance with previous works~\cite{BYS1986,INO04,INO05,BBL05}. These features become more intense when the fluence is larger, as observed in panel (d), which would correspond to a transient regime between the TPP and the thermal PL. Finally, in panel (e) when the regime clearly enters in the thermal PL, all the points in the band structure contribute to smear out any resolvable features. 

Specific features in the electronic band structure that influence NPL depend on the crystal orientation of the material and on the number of atomic planes that compose the thin film. The energy levels and transition dipole matrix elements vary from the quasi-two-dimensional limit of a few MLs to the bulk regime (see Fig.~\ref{FigS5} in Supplementary Figures), and consequently, the resulting NPL changes accordingly. In Fig.~\ref{Fig4}f, we plot the normalized NPL intensity for a set of fluences ranging four orders of magnitude and thus spanning the TPP and thermal PL regimes in few-atom-thick films. In the low-intensity regime, there is a well-defined peak at $\lambda \approx 690$\,nm with an initially thickness-dependent amplitude that converges to a bulk value for a larger number of MLs. As observed in Fig.~\ref{Fig4}c--e, the distinguishable features in reciprocal space fade with increasing pulse fluence, so that a broad range of transitions spanning a larger bandwidth of emission wavelengths can contribute to the overall NPL. Such behavior is reminiscent of blackbody radiation. Besides the general trend regarding the intensity of the impinging electric field, there are spectral differences depending on the number of MLs, which correlate with the band structure and are stronger in the TPP regime.

% ---------------------------------------------------------
\subsection{Experimental inside and theory-experiment comparison}

\begin{figure}
    \centering
    \includegraphics[width=0.45\textwidth]{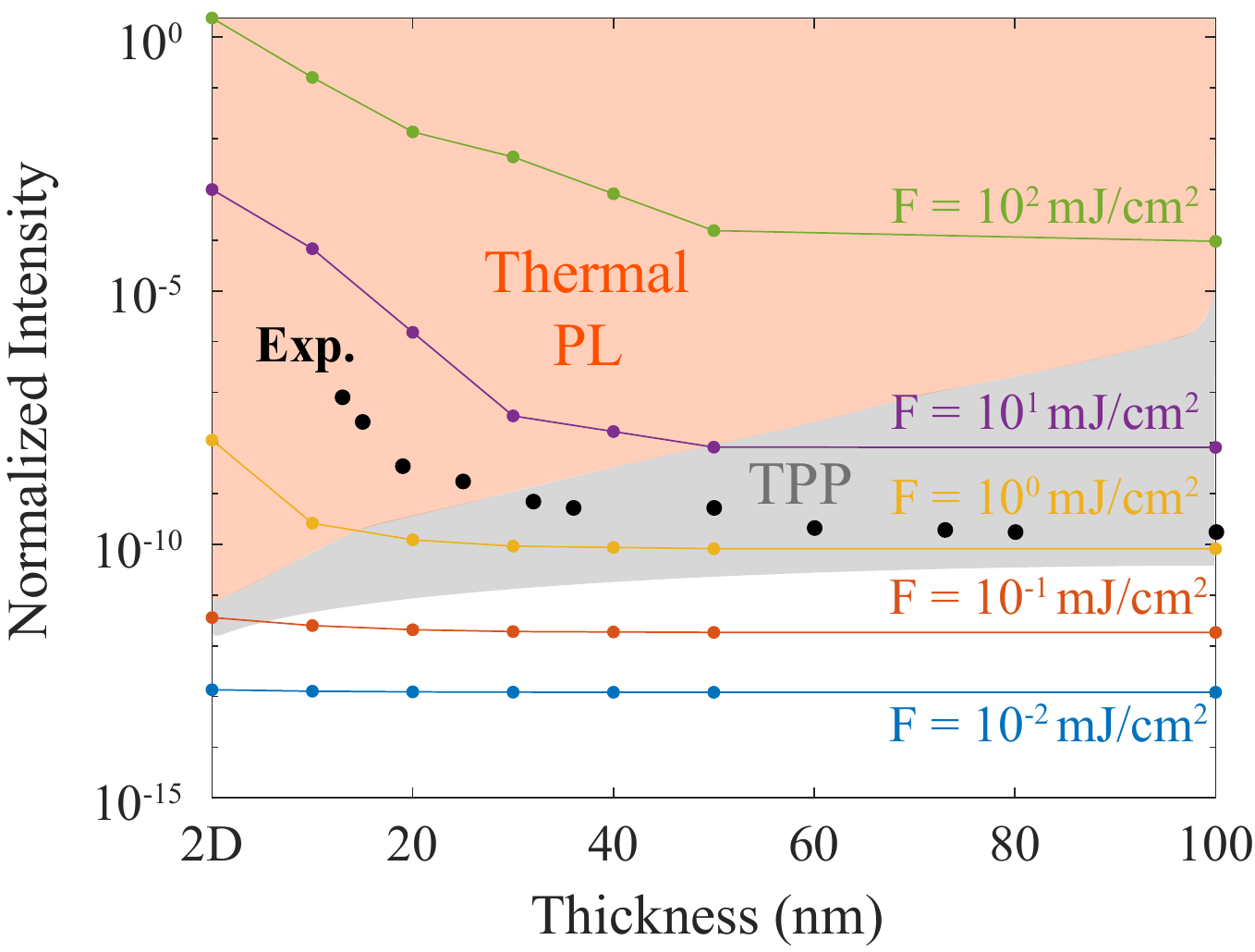}
    \caption{{\bf Photoluminescence as a function of gold film thickness.} We consider different values of the fluence spanning four orders of magnitude (see labels) for a 150\,fs incident Gaussian pulse of 800\,nm central wavelength. Shaded areas indicate the regions where either thermal PL (orange) or TPP (gray) dominates. The white area corresponds to the linear absorption regime. Black dots are experimental values taken from Fig.~\ref{Fig3}a in ref~\citenum{GFK19} for a 120\,fs pulse of 816\,nm central wavelength and $\sim1$\,mJ/cm$^2$ fluence impinging on gold films.}
    \label{Fig5}
\end{figure}

Before analyzing our measurements for crystalline gold flakes, we compare the above theory with the experiments in ref~\citenum{GFK19}, which reports NPL emission from (111) surfaces of monocrystalline gold films with varying thickness. The results are shown in Fig.~\ref{Fig5}, where we plot the thickness dependence of the NPL signal as a function of film thickness for various values of the pump fluence. The superimposed experimental data, normalized to the theory values for a 100\,nm thick film, matches reasonably well with our theory. Furthermore, the transition from the TPP to the thermal PL regime as thickness decreases corresponds to the threshold for which a deviation from the $E^4$ rule is observed in experiment \cite{GFK19}.

% ---------------------------------------------------------

\begin{figure}
    \centering
    \includegraphics[width=0.45\textwidth]{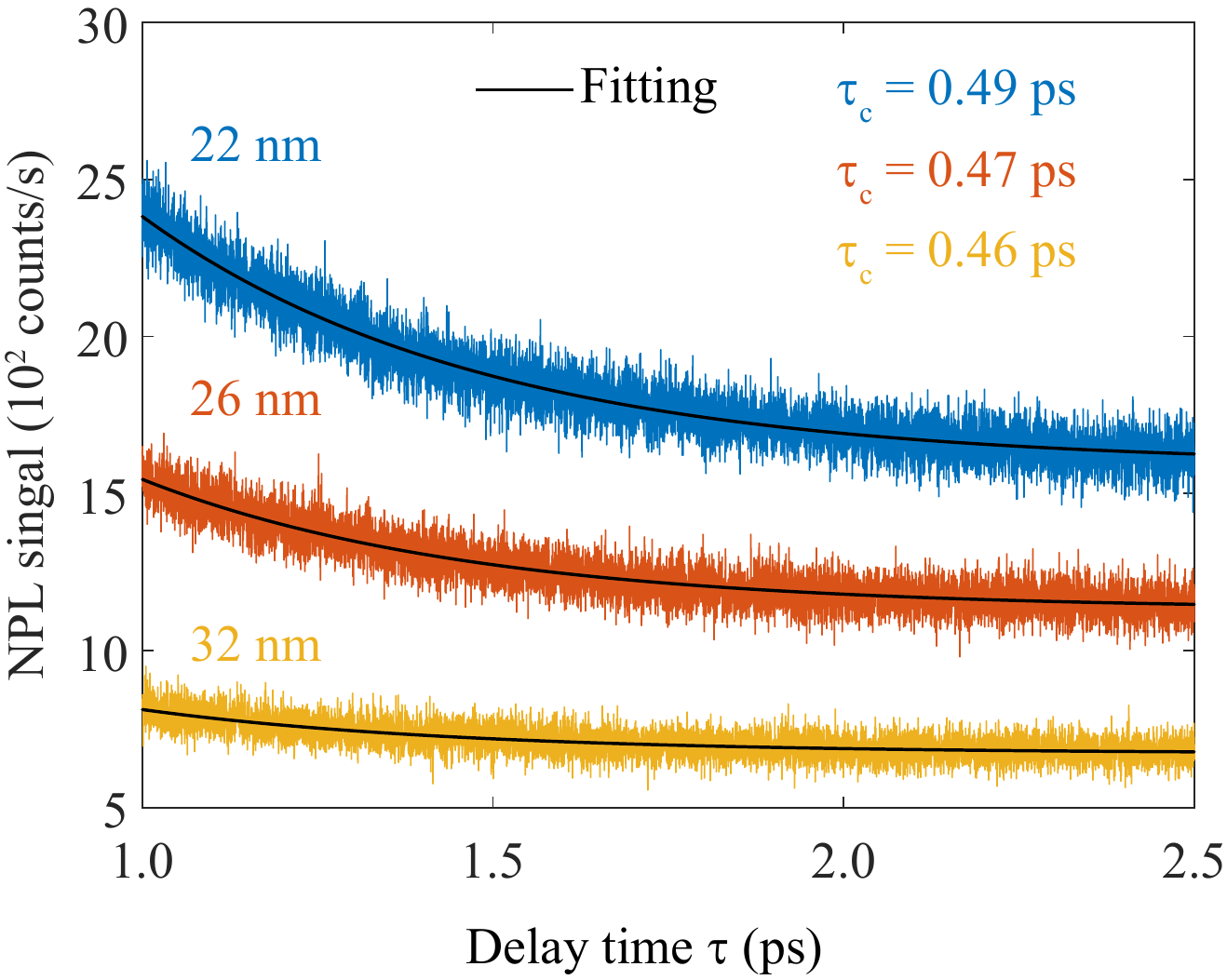}
    \caption{{\bf Two-pulse correlation measurements of PL in thin gold films.} Experimentally obtained two-pulse correlation traces of the NPL signal from the (111) surface of monocrystalline gold films with approximately 22\,nm (blue solid curve), 26\,nm (red), and 32\,nm (yellow) thickness, along with their fitting to $\propto \ee^{-\tau/\tau_{\rm c}}$.} 
    \label{Fig6}
\end{figure}

To obtain additional insight into the NPL dynamics, we perform two-pulse correlation experiments on a monocrystalline gold flake sample that has varying thickness (similar to the samples in ref~\citenum{GFK19}). Further details about the sample preparation and optical characterization can be found in the Appendix. The measured two-pulse correlation traces of the NPL signal for gold thicknesses of approximately 22~nm, 26~nm, and 32~nm obtained at $\sim10$~mJ/cm$^2$ fluence are shown in Fig.~\ref{Fig6}. For the fitting, we consider only the signal obtained for delay times significantly larger than the pulse length ($\tau>1$~ps) to avoid capturing any optical interaction effects (full measurement range correlation curves, including the pulse interference, are provided in Fig.~\ref{FigS6}f in Supplementary Figures). The fitting of transient signals in Fig.~\ref{Fig6} reveals that the characteristic decay time $\tau_{\rm c}$ increases for smaller thickness (from $\sim0.49$~ps to $\sim0.45$~ps). These extracted decay times are comparable to those found in previously reported measurements on thick polycrystalline gold films~\cite{WRL94}, although they are smaller by approximately a factor of 2 in comparison with prior time-resolved NPL measurements on nanostructured gold~\cite{JPJ13,BBH12,BMA15,SBC16,MVP16,XLG19}, in which the reported values are of the order of $\sim1$~ps. Such a discrepancy can be explained by the field enhancement associated with localized surface-plasmon resonances in measurements involving nanostructures or nanoantennas, which give rise to higher electronic temperatures that bring the PL signal to the thermal regime for smaller fluences than for the thin films. In the aforementioned references, the measured decay times are interpreted as the lifetimes of electron-hole pair excitations in the three-step model (i.e., an intermediate state in a cascaded absorption process). According to our model, the observed decay in two-pulse NPL correlation mainly stems from the thermal PL effects provided that the optical interaction time is much smaller (i.e., $\tau_{\rm th} > \tau_{\rm p}$). However, for the experimentally investigated range of thicknesses and fluence, our model predicts longer characteristic decay time $\tau_{\rm c}$ (see Fig.~\ref{Fig3}g), being in a seemingly better accordance with the values reported for nanostructured metals. However, the gold flake sample has a larger lateral dimension ($>10$~$\mu$m) and the excitation spot area is limited to roughly 1~$\mu$m$^2$. This suggests that a lateral diffusion mechanism might take place in the direction parallel to the film surface and, therefore, effectively reduce $\tau_{\rm c}$ as compared to metal nanostructures with dimensions smaller than the excitation spot. Consequently, the experimentally measured decay time can be interpreted as a result of the interplay between the thermal ($\tau_{\rm th}\gtrsim 1$\,ps) and lateral diffusion ($\tau_{\rm d}\gtrsim 1$\,ps) effects, which have comparable (and thus hardly distinguishable) characteristic decay times, as shown in the Appendix. 

% ---------------------------------------------------------

% =========================================================
% --- conclusions -----------------------------------------
% =========================================================
\section{CONCLUSIONS}

We reveal the pivotal role played by electronic thermal effects in the nonlinear optical response (specifically, the nonlinear photoluminescence) of thin metal films depending on thickness and light intensity. We support this conclusion with a theoretical formalism incorporating first-principles details of the electronic band structure, in excellent agreement with available experiments, including our nonlinear photoluminescence measurements for crystalline gold films of 10--100's\,nm thickness.

When the film thickness decreases, the electron temperature at the surface during optical pumping increases considerably, by up to thousands of degrees, making thermal effects dominant under the right combination of incident fluence and thickness. However, thick films are capable of quickly diffusing the absorbed heat into the bulk, thus reducing the influence of thermal effects. Therefore, in the thick-film regime and under relatively low pump intensities (fluences of the order of 50\,mJ/cm$^2$), we conclude that thermal effects are negligible, and TPP dominates the NPL emission with a well-defined fourth power dependence on the amplitude of the incident electric field ($E^4$ rule, or equivalently, the square of the fluence). In contrast, thinner films absorb larger power densities, yielding higher peak temperatures at the surface (i.e., the regions that produce the bulk of the NPL emission, essentially limited to the skin depth) for longer times, and therefore making thermal PL a dominant mechanism that is characterized by an efficiency larger than the $E^4$ rule characteristic of TPP.

Additionally, the TPP and thermal PL mechanisms exhibit distinct characteristic times that we have interrogated by simulating two-pulse correlation of TPP. Exploring the effect of thickness and fluence in the TPP region, the dynamics associated with  two-pulse illumination is controlled by the pulse duration, which in this case is of the order of 150\,fs, whereas for thermal PL the characteristic times are found to be of the order of picoseconds, in agreement with available experiments. Our explanation of such experiments is thus introducing thermal PL as a new ingredient whose importance has been previously overlooked. Additionally, lateral thermal diffusion becomes relevant for extended films, yielding smaller characteristic decay rates compared to finite systems such as nanoparticles, where diffusion is limited.

Further analysis is still required to understand NPL emission from few-atom-thick gold films. We expect that the changes associated with spatial confinement in the band structure will manifest in the TPP signal for metal films of thickness below 10 MLs ($\sim2.3$\,nm), which is a regime accessible to state-of-the-art sample preparation~\cite{paper335}. We envision that the use of atomically thin films with crystalline surface quality will be crucial in the next generation of nanoscale photonic devices, and therefore, the understanding of their optical properties constitutes a current challenge demanding additional explorations in the line suggested by the present work.

% --- acknowledgments -------------------------------------
\section*{ACKNOWLEDGEMENTS}
We thank C. Wolff for stimulating discussions. This work has been supported in part by ERC (Advanced Grant 789104-eNANO), the Spanish MICINN (PID2020-112625-GB-I00 and Severo Ochoa CEX2019-000910-S), the Catalan CERCA Program, and Fundaci\'os Cellex and Mir-Puig. A.R.E. acknowledges support from the Generalitat de Catalunya, the European Social Fund (L\text{'}FSE inverteix en el teu futur)-FEDER. J.D.C. is a Sapere Aude research leader supported by Independent Research Fund Denmark (Grant No.~0165-00051B). N.A.M. is a VILLUM Investigator supported by VILLUM FONDEN (Grant No.~16498). The Center for Polariton-driven Light--Matter Interactions (POLIMA) is funded by the Danish National Research Foundation (Project No.~DNRF165).

% =========================================================
% --- methods/appendix ------------------------------------
% =========================================================
\begin{widetext}
\section*{APPENDIX}
\appendix

% ---------------------------------------------------------
\section{Electronic properties of metal films}

We characterize the electronic structure of a film by calculating the one-electron energies $\energy_{i,\kb}$ as a function of wave vector $\kb$ within the Brillouin zone for each band index $i$ (see the details in below), from which we compute the DOS (see Fig.~\ref{FigS1}a in Supplementary Figures) according to
\begin{align} \label{Eq:DOS}
    \rho^{\rm 3D}(\energy) = \frac{1}{V}\sum_{i,\kb} g_\kb \Gamma(\energy-\energy_{i,\kb}),
\end{align}
where $V$ is the unit-cell volume, $g_\kb$ denotes the weight of each $\kb$ point (here set to $\sum_\kb g_\kb=2$ to account for spin degeneracy), and the function
\begin{align} \label{Eq:smearing}
    \Gamma(\energy) = \frac{1}{\sqrt{2\pi} \hbar\sigma} \ee^{-\energy^2/2(\hbar\sigma)^2}
\end{align}
introduces a phenomenological linewidth that we set to $\hbar\sigma = 0.1$\,eV in order to avoid artefacts associated with the discretized $\kb$-space. The density of occupied states is then defined as $\Omega (\energy) = \rho^{\rm 3D}(\energy) f(\energy,T)$, where the FD distribution at temperature $T$ is computed from eq~\eqref{Eq:FD}, which involves the chemical potential $\mu(T)$ (see Fig.~\ref{FigS1}b in Supplementary Figures). Finite temperature effects are incorporated by computing $\mu(T)$ self-consistently, imposing the conservation of electron density~\cite{AM1976}
\begin{align}
    n_e^{\rm 3D} = \int_{-\infty}^\infty d\energy \,  f(\energy,T) \rho^{\rm 3D}(\energy).
\end{align}
We also require the energy density
\begin{align}
    u(T) = \int_{-\infty}^\infty d\energy \, \energy \, f(\energy,T) \rho^{\rm 3D}(\energy),
\end{align}
from which we obtain the specific heat (Fig.~\ref{FigS1}c in Supplementary Figures) as
\begin{align} \label{Eq:Ce_Au}
    C_e(T) = \left( \frac{\partial u }{ \partial T} \right)_V .
\end{align}

% ---------------------------------------------------------
\section{Thermo-optical response}

We consider an extended thin metal film illuminated by a well-collimated ultrafast optical pulse impinging normal to the film along the $z$ direction, such that the electric field within the film can be approximately written as
\begin{align} \label{Epump1}
    \Eb(z,t) = \Eb_0 A(z)\sqrt{\mathcal{G}(t)}\ee^{\ii(k_z z-\omega_\inn t)} + {\rm c.c.},  
\end{align}
where $\ww_\inn$ is the pulse carrier frequency, the function $A(z)$ accounts for the spatial profile produced by scattering at the film surfaces, and
\begin{align} \label{Eq:Gt_time_profile}
    \mathcal{G}(t) = \sqrt{\frac{4\,\log(2)}{\pi}} \, \text{exp}[-4\,\log(2) \left(\frac{t-2\Delta}{\Delta}\right)^2]
\end{align}
characterizes the temporal Gaussian envelope of the pulse, with a FWHM pulse duration $\Delta$ satisfying the normalization condition $\int_{-\infty}^\infty dt\, \mathcal{G}(t) = \Delta$. Incidentally, we introduce an offset of the pulse center such that the light intensity is negligible at $t=0$. The power absorbed by the film is then given by
\begin{align} \label{Eq:source}
    S(z,t) = I_0 k  {\rm Im} \{ \eps_{\rm m}(\omega)\} \left| A(z) \right|^2 \mathcal{G}(t),
\end{align}
where $\epsilon_{\rm m}(\omega)$ is the bulk dielectric function of the metal and the peak intensity $I_0 = F/\Delta = 2 \pi c |E_0|^2$ is specified in terms of the pulse fluence $F$. For the two-pulse calculations, we superimpose a copy of the original pulse, delayed by a time $\tau$, so that we make the substitution $\mathcal{G}(t) \to \mathcal{G}(t)+\mathcal{G}(t-\tau)$ in eq~\eqref{Eq:source}.

Heating caused by optical pumping produces a raise in the electron temperature $T$ and the lattice temperature $T_l$, which we describe with the TTM given in eq~\eqref{Eq:TTM}, in which $C_e$ is the electronic heat capacity from eq~\eqref{Eq:Ce_Au}, $k_e(T,T_l)=k_0 T/T_l$ is the electronic thermal conductivity in the linear-temperature-dependence regime, with $k_0=317$\,W\,m$^{-1}$K$^{-1}$ for gold~\cite{BDF00}, and $G=2.2 \times 10^{16}$\,W\,m$^{-3}$K$^{-1}$ is the electron-lattice coupling coefficient~\cite{LKL11}. Incidentally, we consider wide pump spots illuminating the samples, so for simplicity, we assume no dependence on lateral position. This assumption is exact for extended plane-wave illumination, but it should be a good approximation for large optical spots compared to the film thickness. Thus, the electron temperature $T(z,t)$ is governed by the one-dimensional diffusion problem of eq~\eqref{Eq:TTM}, which is further simplified by assuming a constant lattice temperature $T_l(t)\approx T_0=293$\,K. This approximation provides a simplified description of the lattice electron cooling channel, and is justified for the pulse fluences considered in this work. The dependence of the electron temperature on distance to the surface and time is illustrated in Fig.~\ref{FigS2} in Supplementary Figures for the types of pump pulses and thicknesses here considered as obtained by numerically solving eq~\eqref{Eq:TTM}. 

The above calculation involves the optical field profile $A(z)$ across the out-of-plane direction (see eq~\eqref{Epump1}). Such profile is obtained by accounting for refraction and transmission at the film surfaces at $z=0$ and $z=\dm$ as well as propagation inside the film of thickness $\dm$. The metal is described by a dielectric function $\epsm(\omega)$ and the film is hosted in a homogeneous dielectric medium of permittivity $\epsd=1$. We approximate the profile by computing it at the central wavelength, such that the optical field consists of plane waves that propagate in the $z$ direction with wave vectors $k_j=\omega \sqrt{\eps_j}/c$ in the metal ($j={\rm m}$) and dielectric ($j={\rm d}$) media. We consider normally incident light, and therefore, the electric field is along an in-plane direction. Then, solving the electromagnetic boundary conditions, we find that the spatial profile of the electric field reads
\begin{align}
    A(z) = 
    \left\{
    \begin{matrix}
    \ee^{\ii k_{\rm d} z} + R  \ee^{-\ii k_{\rm d} z},         &   z<0, \\
    T'  \ee^{\ii k_{\rm m} z} + R' \ee^{-\ii k_{\rm m} z}, & 0<z<\dm, \\
    T \ee^{\ii k_{\rm d} z},                       &   \dm<z,
    \end{matrix}
    \right.
\end{align}
where the film external reflection and transmission coefficients $R$ and $T$, and their respective counterparts $R'$ and $T'$ within the film, are defined by
\begin{subequations}
\begin{align}
    R =& \frac{(\ee^{\ii 2 k_{\rm m} \dm }-1) \rdm }{1-\rmd^2 \ee^{\ii 2 k_{\rm m} \dm }}, &
    T =& \frac{\tdm \tmd \ee^{-\ii (k_{\rm d}-k_{\rm m}) \dm }}{1-\rmd^2 \ee^{\ii 2 k_{\rm m} \dm }}, \\
    R'=& \frac{\tdm \rmd \ee^{\ii 2 k_{\rm m} \dm }}{1-\rmd^2 \ee^{\ii 2 k_{\rm m} \dm }},  &
    T'=& \frac{\tdm }{1-\rmd^2 \ee^{\ii 2 k_{\rm m} \dm }},
\end{align}
\end{subequations}
and expressed in terms of the single interface reflection and transmission coefficients $r_{\rm ab} = (\sqrt{\eps_{\rm b}}-\sqrt{\eps_{\rm a }})/(\sqrt{\eps_{\rm b}}+\sqrt{\eps_{\rm a }})$ and $t_{\rm ab} = 2\sqrt{\eps_{\rm a }}/(\sqrt{\eps_{\rm b}}+\sqrt{\eps_{\rm a }})$ for $\{\rm a,b\} = \{m,d\}$. Here, the temperature-dependent dielectric function of the metal is assumed to be~\cite{paper330}
\begin{align}
    \epsm(\omega,T,T_l) = \eps_{\infty}- \frac{\wp^2}{\ww \left( \omega +\ii \gamma^T \right)} ,
\end{align}
where the background polarizability $\eps_\infty=9.5$ accounts for screening of interband transitions, $\hbar\wp = 9.06$\,eV is the bulk plasma frequency~\cite{JC1972}, and the phenomenological damping rate $\gamma^T \equiv \gamma (T,T_l) = \gamma^{\rm e-e} (T)+\gamma^{\rm e-ph} (T_l)$ is comprised of temperature-dependent electron-electron and electron-phonon contributions that are defined as~\cite{L1976,BC1977,OE17}
 \begin{align} \label{Eq:dampings}
\gamma^{\rm e-e} (T)     = \frac{\pi^3 \alpha \beta}{12\hbar \EF} k_{\rm B}^2T^2
\quad \text{and} \quad
\gamma^{\rm e-ph} (T_l)  \approx \gamma_0 \frac{T_l}{\theta_{\rm D}}
\end{align}
in terms of scattering coefficients $\alpha$ and $\beta$. For gold, we take scattering parameters $\alpha=0.55$ and $\beta=0.77$~\cite{L1976}, a Fermi energy $\EF= 5.53$\,eV~\cite{AM1976}, a Debye temperature $\theta_{\rm D} = 170$\,K~\cite{OE17} equivalent to $\kB\theta_{\rm D}\simeq 15$\,meV, and an inelastic broadening $\hbar \gamma_0 = 39.5$\,eV, such that $\hbar\gamma(T_0,T_0) \approx 71$\,meV at room temperature, in agreement with measured optical data~\cite{JC1972}.

% ---------------------------------------------------------
\section{Electron temperature evolution in the thermalised regime}

Thermal effects have a much longer time scale (of the order of picoseconds, as seen for $\tau_{\rm th}$ in Fig.~\ref{FigS4} in Supplementary Figures) than the optical pump region that extends up to $\sim 300$\,fs after the peak, which corresponds to twice the pulse FWHM. Therefore, after the pump is gone, the temperature quickly thermalises across the film (see Fig.~\ref{FigS2} in Supplementary Figures) and eq~\eqref{Eq:TTM} for the TTM becomes
\begin{align}
    C_e(T) \, \partial_t T = - G(T-T_l).
\end{align}
For temperatures $\lesssim 5\times 10^3$\,K, the heat capacity depends linearly on the temperature as $C_e(T)\approx \gamma T$ (see Fig.~\ref{FigS1}c in Supplementary Figures), in which case the solution of the above differential equation can be obtained analytically as
\begin{align}
    \frac{G}{\gamma} t = (T_i-T) + T_l \,  \text{ln} \left( \frac{T_i-T_l}{T-T_l} \right),
\end{align}
which, for short times when the temperature $T(t)$ is similar to the initial temperature $T_i$, leads to a linear time dependence $ T(t) \approx T_i-(G/\gamma) t$. Such linear behavior can be appreciated in Fig.~\ref{FigS2}c,f,i in Supplementary Figures

% ---------------------------------------------------------
\section{Lateral diffusion}

In the case of thin films (thickness below $\sim 50$\,nm), diffusion in the perpendicular direction to the surface occurs rapidly (see Fig.~\ref{FigS2} in Supplementary Figures), and therefore, the diffusion equation considering a homogeneous excitation (e.g., excitation Gaussian beam) retaining lateral thermal flux terms yields
\begin{align}
    C_e(T) \, \partial_t T = R^{-1} \partial_R \left[R \, k_e(T) \partial_R T\right] - G(T-T_l),
\end{align}
in cylindrical coordinates, where $R$ is the radial coordinate starting from the center of the excitation source. If we only consider the dispersion part (i.e., neglecting the coupling to the lattice term $G(T-T_l)$), and assuming that we are in the linear regime of the heat capacity $C_e(T)\approx \gamma T$ (see Fig.~\ref{FigS1}c in Supplementary Figures), the above equation is satisfied by 
\begin{align}
    T(R,t) =  \frac{ T_i \sigma }{\sqrt{\sigma^2 + 8(k/\gamma)t } } \ee^{- \frac{R^2}{\sigma^2 + 8(k/\gamma)t}}
\end{align}
provided a spatially Gaussian distribution as an initial condition, which corresponds to a Gaussian that keeps the integrated area constant at any time. Interestingly, near the center of the Gaussian (i.e., $R \approx 0$), the time for which its temperature drops $1/\ee$ from the initial value $T_i$ results in $ \tau_{\rm d} =   \sigma^2  (\ee^2-1) \gamma/8 k $, which for a beam waist $\sigma \sim \lambda/2  = 400$\,nm, $k\equiv k_0/T_l = 1.08 $\,Wm$^{-1}$, and $\gamma \approx 70$\,Jm$^{-3}$K$^{-2}$, give rise to $\tau_{\rm d} \approx 8$\,ps.

% ---------------------------------------------------------
\section{Electronic dynamics}

The electronic distribution of the metal film is described by the occupation factors $0\le p_i(\kb)\le1$ of band $i$ at wave vector $\kb$, which are initially determined by the FD distribution of eq~\eqref{Eq:FD} at temperature $T_0$ (i.e., $p^{(0)}_{i}(\kb) = f(\energy_{i,\kb},T_0)$). The time evolution of the electronic distribution during optical excitation and subsequent thermalization is described by the population rate equation given in eq~\eqref{Eq:rate_align}. The electronic wave vector $\kb$ is assumed to be conserved in the transitions associated with light absorption or emission, as the photon momentum is negligible compared to the size of the Brillouin zone at the optical frequencies under consideration. Nevertheless, non-vertical transitions are phenomenologically incorporated through the thermal relaxation term in eq~\eqref{Eq:rate_align}. The optical excitation is described through a rate $\gamma_{ij}^{\rm ex}=\gamma^{(1)}_{ij}+\gamma^{(2)}_{ij}$, comprising a linear absorption component,
\begin{align}
    \gamma^{(1)}_{ij} (t,\omega_\inn) = \frac{2\pi}{\hbar} E_0^2 \Gg(t)
    |\mub_{ij}|^2\sum_\pm 
    \Gamma(\energy_{ij}\pm\hbar \omega_\inn),
\end{align}
and a two-photon absorption part,
\begin{align}
    \gamma^{(2)}_{ij} (t,\omega_\inn) = \frac{2\pi}{\hbar} E_0^4 \Gg^2(t) \sum_\pm \Gamma(\energy_{ij}\pm2\hbar \omega_\inn)    \sum_{a,b=\{x,y\}}\left| \sum_l \frac{\mu_{il,a} \mu_{lj,b}}{\energy_{lj}\pm\hbar\ww_\inn+\ii\hbar \sigma} \right|^2
\end{align}
both expressed in terms of the linewidth of the final state $\Gamma(\energy)$ given in eq~\eqref{Eq:smearing}, the field intensity $|E_0|^2 =I_0/2\pi c$, the temporal pulse profile $\Gg(t)$ in eq~\eqref{Eq:Gt_time_profile}, the energy differences $\energy_{ij}=\energy_{i,\kb}-\energy_{j,\kb}$, and the electronic transition dipole matrix elements $\mub_{ij}\equiv-e\langle i,\kb|\Rb|j,\kb\rangle$~\cite{B08_3}, with in-plane coordinates $\Rb = (x,y)$. Note that the two contributions at frequencies $\pm \ww$ in either optical excitation process account for absorption ($+$) and stimulated emission ($-$) that excite or de-excite electrons when the population difference $p_j-p_i$ is nonzero according to eq~\eqref{Eq:rate_align}. Spontaneous emission acts only to de-excite electrons to lower energies at a rate 
\begin{align}
    \gamma^\sp_{ij} = \frac{4e^2 \energy_{ji}^3}{3\hbar^4} 
    |\mub_{ij}|^2
    \Theta(\energy_{j}-\energy_{i}),
\end{align}
which is directly related to the Einstein $A$-coefficient~\cite{ST19}.

% ---------------------------------------------------------
\section{Density-functional theory (DFT) calculations}

We obtain the one-electron band structure (e.g., see Fig.~\ref{FigS5} in Supplementary Figures) and dipole transition matrix elements of thin gold films consisting of up to 50 MLs from density-functional theory (DFT) calculations using Quantum Espresso~\cite{GBB09} combined with the Perdew--Burke--Ernzerhof (PBE)~\cite{PBE96} parametrization of the generalized gradient approximation. We adopt optimized norm-conserving Vanderbilt~\cite{H13_3,SG15} pseudopotentials with a kinetic energy cut-off of 80\,Ry. Energy minimization yields a bulk lattice constant of 4.155\,{\AA}, which we use to construct thin gold films by fixing the interatomic bond distances. A vacuum spacing of 10\,{\AA} is introduced in the vertical direction to avoid spurious interactions between neighboring cells. We use a 24$\times$24$\times$1$\kb$-grid to calculate electronic structures. Using the resulting Kohn--Sham wave functions and eigenvalues, we obtain the transition dipole matrix elements $\mub_{ij}$ using the YAMBO code~\cite{MHG09}.

% ---------------------------------------------------------
\section{Experimental methods}

The gold flake sample was synthesized in an endothermic reduction of HAuC$_{l4}$ precursor, following the prescription of refs~\citenum{KKW18} and \citenum{KT22}. In short, 20~$\mu$l of 0.5~M aqueous solution of HAuC$_{l4}$ was mixed with 5~mm of ethylene glycole (both reagents from purchased from Sigma Aldrich) and vigorously stirred. Further, a pre-cleaned (ultrasonication in acetone, isopropyl alcohol and deionized water) glass cover slip (Menzel 1) was immersed in the solution and left on a hot plate at 100$^\circ$C. After 24~h, the glass cover slip is removed from the growth solution, rinsed with isopropyl alcohol and deionized water, and dried in a nitrogen flow. This procedure results in a large batch of gold flake samples with varying thicknesses. An optical image, atomic-force microscope topography and a NPL confocal map of the investigated flake sample are shown in Fig.~\ref{FigS6}a-c in Supplementary Figures.

Two-pulse correlation measurements were performed using an experimental setup based on a custom-made scanning nonlinear microscope (schematics of the setup are shown in Fig.~\ref{FigS6}g in Supplementary Figures). The laser source used in the experiment is a mode-locked titanium-sapphire (Ti:sapph) oscillator (Tsunami 3941 by Spectra-Physics), which provides  approximately 120~ps pulses with central wavelength 800~nm. The delay between the pulses is controlled by moving a mirror in the delay arm of the Michelson interferometer setup, through which the laser beam goes before entering the microscope. In the two-pulse correlation measurements, we detect only the NPL content of the broad nonlinear signal (see spectral measurements in Fig.~\ref{FigS6}d in Supplementary Figures): the second-harmonic peak is filtered out using an appropriate band-pass filter. The measurements were performed in reflection mode, using a ${\rm NA}=0.9$ objective to focus the excitation light down to a diffraction-limited spot ($\sim1$~$\mu$m diameter) and collect the nonlinear signal. To maximize the signal, we used approximately 10~mJ/cm$^2$ fluence (averaged within the focal spot), which, according to our fluence dependence measurements (see Fig.~\ref{FigS6}e in Supplementary Figures) is still within the two-photon process range.

\end{widetext}

%\bibliography{../../../bibtex/refs}

\begin{thebibliography}{64}%
\makeatletter
\providecommand \@ifxundefined [1]{%
 \@ifx{#1\undefined}
}%
\providecommand \@ifnum [1]{%
 \ifnum #1\expandafter \@firstoftwo
 \else \expandafter \@secondoftwo
 \fi
}%
\providecommand \@ifx [1]{%
 \ifx #1\expandafter \@firstoftwo
 \else \expandafter \@secondoftwo
 \fi
}%
\providecommand \natexlab [1]{#1}%
\providecommand \enquote  [1]{``#1''}%
\providecommand \bibnamefont  [1]{#1}%
\providecommand \bibfnamefont [1]{#1}%
\providecommand \citenamefont [1]{#1}%
\providecommand \href@noop [0]{\@secondoftwo}%
\providecommand \href [0]{\begingroup \@sanitize@url \@href}%
\providecommand \@href[1]{\@@startlink{#1}\@@href}%
\providecommand \@@href[1]{\endgroup#1\@@endlink}%
\providecommand \@sanitize@url [0]{\catcode `\\12\catcode `\$12\catcode
  `\&12\catcode `\#12\catcode `\^12\catcode `\_12\catcode `\%12\relax}%
\providecommand \@@startlink[1]{}%
\providecommand \@@endlink[0]{}%
\providecommand \url  [0]{\begingroup\@sanitize@url \@url }%
\providecommand \@url [1]{\endgroup\@href {#1}{\urlprefix }}%
\providecommand \urlprefix  [0]{URL }%
\providecommand \Eprint [0]{\href }%
\providecommand \doibase [0]{http://dx.doi.org/}%
\providecommand \selectlanguage [0]{\@gobble}%
\providecommand \bibinfo  [0]{\@secondoftwo}%
\providecommand \bibfield  [0]{\@secondoftwo}%
\providecommand \translation [1]{[#1]}%
\providecommand \BibitemOpen [0]{}%
\providecommand \bibitemStop [0]{}%
\providecommand \bibitemNoStop [0]{.\EOS\space}%
\providecommand \EOS [0]{\spacefactor3000\relax}%
\providecommand \BibitemShut  [1]{\csname bibitem#1\endcsname}%
\let\auto@bib@innerbib\@empty
%</preamble>
\bibitem [{\citenamefont {Mooradian}(1969)}]{M1969}%
  \BibitemOpen
  \bibfield  {author} {\bibinfo {author} {\bibfnamefont {A}~\bibnamefont
  {Mooradian}},\ }\bibfield  {title} {\enquote {\bibinfo {title}
  {Photoluminescence of metals},}\ }\href@noop {} {\bibfield  {journal}
  {\bibinfo  {journal} {Phys.\ Rev.\ Lett.}\ }\textbf {\bibinfo {volume}
  {22}},\ \bibinfo {pages} {185} (\bibinfo {year} {1969})}\BibitemShut
  {NoStop}%
\bibitem [{\citenamefont {Farrer}\ \emph {et~al.}(2005)\citenamefont {Farrer},
  \citenamefont {Butterfield}, \citenamefont {Chen},\ and\ \citenamefont
  {Fourkas}}]{FBC05}%
  \BibitemOpen
  \bibfield  {author} {\bibinfo {author} {\bibfnamefont {Richard~A.}\
  \bibnamefont {Farrer}}, \bibinfo {author} {\bibfnamefont {Francis~L.}\
  \bibnamefont {Butterfield}}, \bibinfo {author} {\bibfnamefont {Vincent~W.}\
  \bibnamefont {Chen}}, \ and\ \bibinfo {author} {\bibfnamefont {John~T.}\
  \bibnamefont {Fourkas}},\ }\bibfield  {title} {\enquote {\bibinfo {title}
  {Highly efficient multiphoton-absorption-induced luminescence from gold
  nanoparticles},}\ }\href@noop {} {\bibfield  {journal} {\bibinfo  {journal}
  {Nano\ Lett.}\ }\textbf {\bibinfo {volume} {5}},\ \bibinfo {pages}
  {1139--1142} (\bibinfo {year} {2005})}\BibitemShut {NoStop}%
\bibitem [{\citenamefont {Beversluis}\ \emph {et~al.}(2003)\citenamefont
  {Beversluis}, \citenamefont {Bouhelier},\ and\ \citenamefont
  {Novotny}}]{BBN03_1}%
  \BibitemOpen
  \bibfield  {author} {\bibinfo {author} {\bibfnamefont {Michael~R}\
  \bibnamefont {Beversluis}}, \bibinfo {author} {\bibfnamefont {Alexandre}\
  \bibnamefont {Bouhelier}}, \ and\ \bibinfo {author} {\bibfnamefont {Lukas}\
  \bibnamefont {Novotny}},\ }\bibfield  {title} {\enquote {\bibinfo {title}
  {Continuum generation from single gold nanostructures through near-field
  mediated intraband transitions},}\ }\href {\doibase
  10.1103/PhysRevB.68.115433} {\bibfield  {journal} {\bibinfo  {journal}
  {Phys.\ Rev.\ B}\ }\textbf {\bibinfo {volume} {68}},\ \bibinfo {pages}
  {115433} (\bibinfo {year} {2003})}\BibitemShut {NoStop}%
\bibitem [{\citenamefont {Imura}\ \emph {et~al.}(2004)\citenamefont {Imura},
  \citenamefont {Nagahara},\ and\ \citenamefont {Okamoto}}]{INO04}%
  \BibitemOpen
  \bibfield  {author} {\bibinfo {author} {\bibfnamefont {Kohei}\ \bibnamefont
  {Imura}}, \bibinfo {author} {\bibfnamefont {Tetsuhiko}\ \bibnamefont
  {Nagahara}}, \ and\ \bibinfo {author} {\bibfnamefont {Hiromi}\ \bibnamefont
  {Okamoto}},\ }\bibfield  {title} {\enquote {\bibinfo {title} {Plasmon mode
  imaging of single gold nanorods},}\ }\href@noop {} {\bibfield  {journal}
  {\bibinfo  {journal} {J.\ Am.\ Chem.\ Soc.}\ }\textbf {\bibinfo {volume}
  {126}},\ \bibinfo {pages} {12730--12731} (\bibinfo {year}
  {2004})}\BibitemShut {NoStop}%
\bibitem [{\citenamefont {Yelin}\ \emph {et~al.}(2003)\citenamefont {Yelin},
  \citenamefont {Oron}, \citenamefont {Thiberge}, \citenamefont {Moses},\ and\
  \citenamefont {Silberberg}}]{YOT03}%
  \BibitemOpen
  \bibfield  {author} {\bibinfo {author} {\bibfnamefont {Dvir}\ \bibnamefont
  {Yelin}}, \bibinfo {author} {\bibfnamefont {Dan}\ \bibnamefont {Oron}},
  \bibinfo {author} {\bibfnamefont {Stephan}\ \bibnamefont {Thiberge}},
  \bibinfo {author} {\bibfnamefont {Elisha}\ \bibnamefont {Moses}}, \ and\
  \bibinfo {author} {\bibfnamefont {Yaron}\ \bibnamefont {Silberberg}},\
  }\bibfield  {title} {\enquote {\bibinfo {title} {Multiphoton
  plasmon-resonance microscopy},}\ }\href@noop {} {\bibfield  {journal}
  {\bibinfo  {journal} {Opt.\ Express}\ }\textbf {\bibinfo {volume} {11}},\
  \bibinfo {pages} {1385--1391} (\bibinfo {year} {2003})}\BibitemShut {NoStop}%
\bibitem [{\citenamefont {Sheppard}(2020)}]{S20}%
  \BibitemOpen
  \bibfield  {author} {\bibinfo {author} {\bibfnamefont {Colin~JR}\
  \bibnamefont {Sheppard}},\ }\bibfield  {title} {\enquote {\bibinfo {title}
  {Multiphoton microscopy: a personal historical review, with some future
  predictions},}\ }\href@noop {} {\bibfield  {journal} {\bibinfo  {journal}
  {J.\ Biomed.\ Opt.}\ }\textbf {\bibinfo {volume} {25}},\ \bibinfo {pages}
  {014511} (\bibinfo {year} {2020})}\BibitemShut {NoStop}%
\bibitem [{\citenamefont {Petek}\ and\ \citenamefont {Ogawa}(1997)}]{PO97}%
  \BibitemOpen
  \bibfield  {author} {\bibinfo {author} {\bibfnamefont {H.}~\bibnamefont
  {Petek}}\ and\ \bibinfo {author} {\bibfnamefont {S.}~\bibnamefont {Ogawa}},\
  }\bibfield  {title} {\enquote {\bibinfo {title} {Femtosecond time-resolved
  two-photon photoemission studies of electron dynamics in metals},}\
  }\href@noop {} {\bibfield  {journal} {\bibinfo  {journal} {Prog.\ Surf.\
  Sci.}\ }\textbf {\bibinfo {volume} {56}},\ \bibinfo {pages} {239--310}
  (\bibinfo {year} {1997})}\BibitemShut {NoStop}%
\bibitem [{\citenamefont {{Abd El-Fattah}}\ \emph {et~al.}(2019)\citenamefont
  {{Abd El-Fattah}}, \citenamefont {Mkhitaryan}, \citenamefont {Brede},
  \citenamefont {Fern\'andez}, \citenamefont {Li}, \citenamefont {Guo},
  \citenamefont {Ghosh}, \citenamefont {{Rodr\'{\i}guez Echarri}},
  \citenamefont {Naveh}, \citenamefont {Xia}, \citenamefont {Ortega},\ and\
  \citenamefont {{Garc\'{\i}a de Abajo}}}]{paper335}%
  \BibitemOpen
  \bibfield  {author} {\bibinfo {author} {\bibfnamefont {Z.~M.}\ \bibnamefont
  {{Abd El-Fattah}}}, \bibinfo {author} {\bibfnamefont {V.}~\bibnamefont
  {Mkhitaryan}}, \bibinfo {author} {\bibfnamefont {J.}~\bibnamefont {Brede}},
  \bibinfo {author} {\bibfnamefont {L.}~\bibnamefont {Fern\'andez}}, \bibinfo
  {author} {\bibfnamefont {C.}~\bibnamefont {Li}}, \bibinfo {author}
  {\bibfnamefont {Q.}~\bibnamefont {Guo}}, \bibinfo {author} {\bibfnamefont
  {A.}~\bibnamefont {Ghosh}}, \bibinfo {author} {\bibfnamefont
  {A.}~\bibnamefont {{Rodr\'{\i}guez Echarri}}}, \bibinfo {author}
  {\bibfnamefont {D.}~\bibnamefont {Naveh}}, \bibinfo {author} {\bibfnamefont
  {F.}~\bibnamefont {Xia}}, \bibinfo {author} {\bibfnamefont {J.~E.}\
  \bibnamefont {Ortega}}, \ and\ \bibinfo {author} {\bibfnamefont {F.~J.}\
  \bibnamefont {{Garc\'{\i}a de Abajo}}},\ }\bibfield  {title} {\enquote
  {\bibinfo {title} {Plasmonics in atomically thin crystalline silver films},}\
  }\href {\doibase 10.1021/acsnano.9b01651} {\bibfield  {journal} {\bibinfo
  {journal} {ACS\ Nano}\ }\textbf {\bibinfo {volume} {13}},\ \bibinfo {pages}
  {7771--7779} (\bibinfo {year} {2019})}\BibitemShut {NoStop}%
\bibitem [{\citenamefont {Dryzek}\ and\ \citenamefont {Czapla}(1987)}]{DC1987}%
  \BibitemOpen
  \bibfield  {author} {\bibinfo {author} {\bibfnamefont {J.}~\bibnamefont
  {Dryzek}}\ and\ \bibinfo {author} {\bibfnamefont {A.}~\bibnamefont
  {Czapla}},\ }\bibfield  {title} {\enquote {\bibinfo {title} {Quantum size
  effect in optical spectra of thin metallic films},}\ }\href@noop {}
  {\bibfield  {journal} {\bibinfo  {journal} {Phys.\ Rev.\ Lett.}\ }\textbf
  {\bibinfo {volume} {58}},\ \bibinfo {pages} {721} (\bibinfo {year}
  {1987})}\BibitemShut {NoStop}%
\bibitem [{\citenamefont {Qian}\ \emph {et~al.}(2015)\citenamefont {Qian},
  \citenamefont {Xiao}, \citenamefont {Lepage}, \citenamefont {Chen},\ and\
  \citenamefont {Liu}}]{QXL15}%
  \BibitemOpen
  \bibfield  {author} {\bibinfo {author} {\bibfnamefont {Haoliang}\
  \bibnamefont {Qian}}, \bibinfo {author} {\bibfnamefont {Yuzhe}\ \bibnamefont
  {Xiao}}, \bibinfo {author} {\bibfnamefont {Dominic}\ \bibnamefont {Lepage}},
  \bibinfo {author} {\bibfnamefont {Li}~\bibnamefont {Chen}}, \ and\ \bibinfo
  {author} {\bibfnamefont {Zhaowei}\ \bibnamefont {Liu}},\ }\bibfield  {title}
  {\enquote {\bibinfo {title} {Quantum electrostatic model for optical
  properties of nanoscale gold films},}\ }\href@noop {} {\bibfield  {journal}
  {\bibinfo  {journal} {Nanophotonics}\ }\textbf {\bibinfo {volume} {4}},\
  \bibinfo {pages} {413--418} (\bibinfo {year} {2015})}\BibitemShut {NoStop}%
\bibitem [{\citenamefont {Qian}\ \emph {et~al.}(2016)\citenamefont {Qian},
  \citenamefont {Xiao},\ and\ \citenamefont {Liu}}]{QXL16}%
  \BibitemOpen
  \bibfield  {author} {\bibinfo {author} {\bibfnamefont {Haoliang}\
  \bibnamefont {Qian}}, \bibinfo {author} {\bibfnamefont {Yuzhe}\ \bibnamefont
  {Xiao}}, \ and\ \bibinfo {author} {\bibfnamefont {Zhaowei}\ \bibnamefont
  {Liu}},\ }\bibfield  {title} {\enquote {\bibinfo {title} {Giant kerr response
  of ultrathin gold films from quantum size effect},}\ }\href@noop {}
  {\bibfield  {journal} {\bibinfo  {journal} {Nat.\ Commun.}\ }\textbf
  {\bibinfo {volume} {7}},\ \bibinfo {pages} {13153} (\bibinfo {year}
  {2016})}\BibitemShut {NoStop}%
\bibitem [{\citenamefont {Gro{\ss}mann}\ \emph {et~al.}(2019)\citenamefont
  {Gro{\ss}mann}, \citenamefont {Friedrich}, \citenamefont {Karolak},
  \citenamefont {Kullock}, \citenamefont {Krauss}, \citenamefont {Emmerling},
  \citenamefont {Sangiovanni},\ and\ \citenamefont {Hecht}}]{GFK19}%
  \BibitemOpen
  \bibfield  {author} {\bibinfo {author} {\bibfnamefont {Swen}\ \bibnamefont
  {Gro{\ss}mann}}, \bibinfo {author} {\bibfnamefont {Daniel}\ \bibnamefont
  {Friedrich}}, \bibinfo {author} {\bibfnamefont {Michael}\ \bibnamefont
  {Karolak}}, \bibinfo {author} {\bibfnamefont {Ren{\'e}}\ \bibnamefont
  {Kullock}}, \bibinfo {author} {\bibfnamefont {Enno}\ \bibnamefont {Krauss}},
  \bibinfo {author} {\bibfnamefont {Monika}\ \bibnamefont {Emmerling}},
  \bibinfo {author} {\bibfnamefont {Giorgio}\ \bibnamefont {Sangiovanni}}, \
  and\ \bibinfo {author} {\bibfnamefont {Bert}\ \bibnamefont {Hecht}},\
  }\bibfield  {title} {\enquote {\bibinfo {title} {Nonclassical optical
  properties of mesoscopic gold},}\ }\href@noop {} {\bibfield  {journal}
  {\bibinfo  {journal} {Phys.\ Rev.\ Lett.}\ }\textbf {\bibinfo {volume}
  {122}},\ \bibinfo {pages} {246802} (\bibinfo {year} {2019})}\BibitemShut
  {NoStop}%
\bibitem [{\citenamefont {Chen}\ \emph {et~al.}(1981)\citenamefont {Chen},
  \citenamefont {{De Castro}},\ and\ \citenamefont {Shen}}]{CCS1981}%
  \BibitemOpen
  \bibfield  {author} {\bibinfo {author} {\bibfnamefont {C.K.}\ \bibnamefont
  {Chen}}, \bibinfo {author} {\bibfnamefont {A.R.B.}\ \bibnamefont {{De
  Castro}}}, \ and\ \bibinfo {author} {\bibfnamefont {Y.R.}\ \bibnamefont
  {Shen}},\ }\bibfield  {title} {\enquote {\bibinfo {title} {Surface-enhanced
  second-harmonic generation},}\ }\href@noop {} {\bibfield  {journal} {\bibinfo
   {journal} {Phys.\ Rev.\ Lett.}\ }\textbf {\bibinfo {volume} {46}},\ \bibinfo
  {pages} {145} (\bibinfo {year} {1981})}\BibitemShut {NoStop}%
\bibitem [{\citenamefont {Hubert}\ \emph {et~al.}(2007)\citenamefont {Hubert},
  \citenamefont {Billot}, \citenamefont {Adam}, \citenamefont {Bachelot},
  \citenamefont {Royer}, \citenamefont {Grand}, \citenamefont {Gindre},
  \citenamefont {Dorkenoo},\ and\ \citenamefont {Fort}}]{HBA07}%
  \BibitemOpen
  \bibfield  {author} {\bibinfo {author} {\bibfnamefont {Christophe}\
  \bibnamefont {Hubert}}, \bibinfo {author} {\bibfnamefont {Laurent}\
  \bibnamefont {Billot}}, \bibinfo {author} {\bibfnamefont {P.-M.}\
  \bibnamefont {Adam}}, \bibinfo {author} {\bibfnamefont {Renaud}\ \bibnamefont
  {Bachelot}}, \bibinfo {author} {\bibfnamefont {Pascal}\ \bibnamefont
  {Royer}}, \bibinfo {author} {\bibfnamefont {Johan}\ \bibnamefont {Grand}},
  \bibinfo {author} {\bibfnamefont {Denis}\ \bibnamefont {Gindre}}, \bibinfo
  {author} {\bibfnamefont {K.~D.}\ \bibnamefont {Dorkenoo}}, \ and\ \bibinfo
  {author} {\bibfnamefont {Alain}\ \bibnamefont {Fort}},\ }\bibfield  {title}
  {\enquote {\bibinfo {title} {Role of surface plasmon in second harmonic
  generation from gold nanorods},}\ }\href@noop {} {\bibfield  {journal}
  {\bibinfo  {journal} {Appl.\ Phys.\ Lett.}\ }\textbf {\bibinfo {volume}
  {90}},\ \bibinfo {pages} {181105} (\bibinfo {year} {2007})}\BibitemShut
  {NoStop}%
\bibitem [{\citenamefont {Slablab}\ \emph {et~al.}(2012)\citenamefont
  {Slablab}, \citenamefont {Le~Xuan}, \citenamefont {Zielinski}, \citenamefont
  {De~Wilde}, \citenamefont {Jacques}, \citenamefont {Chauvat},\ and\
  \citenamefont {Roch}}]{SLZ12}%
  \BibitemOpen
  \bibfield  {author} {\bibinfo {author} {\bibfnamefont {A.}~\bibnamefont
  {Slablab}}, \bibinfo {author} {\bibfnamefont {L.}~\bibnamefont {Le~Xuan}},
  \bibinfo {author} {\bibfnamefont {M.}~\bibnamefont {Zielinski}}, \bibinfo
  {author} {\bibfnamefont {Y.}~\bibnamefont {De~Wilde}}, \bibinfo {author}
  {\bibfnamefont {V.}~\bibnamefont {Jacques}}, \bibinfo {author} {\bibfnamefont
  {D.}~\bibnamefont {Chauvat}}, \ and\ \bibinfo {author} {\bibfnamefont
  {J.-F.}\ \bibnamefont {Roch}},\ }\bibfield  {title} {\enquote {\bibinfo
  {title} {Second-harmonic generation from coupled plasmon modes in a single
  dimer of gold nanospheres},}\ }\href@noop {} {\bibfield  {journal} {\bibinfo
  {journal} {Opt.\ Express}\ }\textbf {\bibinfo {volume} {20}},\ \bibinfo
  {pages} {220--227} (\bibinfo {year} {2012})}\BibitemShut {NoStop}%
\bibitem [{\citenamefont {Jiang}\ \emph {et~al.}(2013)\citenamefont {Jiang},
  \citenamefont {Pan}, \citenamefont {Jiang}, \citenamefont {Zhao},
  \citenamefont {Yuan}, \citenamefont {Venkatesan},\ and\ \citenamefont
  {Xu}}]{JPJ13}%
  \BibitemOpen
  \bibfield  {author} {\bibinfo {author} {\bibfnamefont {Xiao-Fang}\
  \bibnamefont {Jiang}}, \bibinfo {author} {\bibfnamefont {Yanlin}\
  \bibnamefont {Pan}}, \bibinfo {author} {\bibfnamefont {Cuifeng}\ \bibnamefont
  {Jiang}}, \bibinfo {author} {\bibfnamefont {Tingting}\ \bibnamefont {Zhao}},
  \bibinfo {author} {\bibfnamefont {Peiyan}\ \bibnamefont {Yuan}}, \bibinfo
  {author} {\bibfnamefont {T.}~\bibnamefont {Venkatesan}}, \ and\ \bibinfo
  {author} {\bibfnamefont {Qing-Hua}\ \bibnamefont {Xu}},\ }\bibfield  {title}
  {\enquote {\bibinfo {title} {Excitation nature of two-photon
  photoluminescence of gold nanorods and coupled gold nanoparticles studied by
  two-pulse emission modulation spectroscopy},}\ }\href@noop {} {\bibfield
  {journal} {\bibinfo  {journal} {J.\ Phys.\ Chem.\ Lett.}\ }\textbf {\bibinfo
  {volume} {4}},\ \bibinfo {pages} {1634--1638} (\bibinfo {year}
  {2013})}\BibitemShut {NoStop}%
\bibitem [{\citenamefont {Xie}\ \emph {et~al.}(2019)\citenamefont {Xie},
  \citenamefont {Laforge}, \citenamefont {Grigorenko},\ and\ \citenamefont
  {Rabitz}}]{XLG19}%
  \BibitemOpen
  \bibfield  {author} {\bibinfo {author} {\bibfnamefont {Dan}\ \bibnamefont
  {Xie}}, \bibinfo {author} {\bibfnamefont {Fran{\c{c}}ois~O.}\ \bibnamefont
  {Laforge}}, \bibinfo {author} {\bibfnamefont {Ilya}\ \bibnamefont
  {Grigorenko}}, \ and\ \bibinfo {author} {\bibfnamefont {Herschel~A.}\
  \bibnamefont {Rabitz}},\ }\bibfield  {title} {\enquote {\bibinfo {title}
  {Dual coherent and incoherent two-photon luminescence in single gold nanorods
  revealed by polarization and time-resolved nonlinear autocorrelation},}\
  }\href@noop {} {\bibfield  {journal} {\bibinfo  {journal} {J.\ Opt.\ Soc.\
  Am.\ B}\ }\textbf {\bibinfo {volume} {36}},\ \bibinfo {pages} {1931--1936}
  (\bibinfo {year} {2019})}\BibitemShut {NoStop}%
\bibitem [{\citenamefont {Boyd}\ \emph {et~al.}(1986)\citenamefont {Boyd},
  \citenamefont {Yu},\ and\ \citenamefont {Shen}}]{BYS1986}%
  \BibitemOpen
  \bibfield  {author} {\bibinfo {author} {\bibfnamefont {G.~T.}\ \bibnamefont
  {Boyd}}, \bibinfo {author} {\bibfnamefont {Z.~H.}\ \bibnamefont {Yu}}, \ and\
  \bibinfo {author} {\bibfnamefont {Y.~R.}\ \bibnamefont {Shen}},\ }\bibfield
  {title} {\enquote {\bibinfo {title} {Photoinduced luminescence from the noble
  metals and its enhancement on roughened surfaces},}\ }\href@noop {}
  {\bibfield  {journal} {\bibinfo  {journal} {Phys.\ Rev.\ B}\ }\textbf
  {\bibinfo {volume} {33}},\ \bibinfo {pages} {7923} (\bibinfo {year}
  {1986})}\BibitemShut {NoStop}%
\bibitem [{\citenamefont {Imura}\ \emph {et~al.}(2005)\citenamefont {Imura},
  \citenamefont {Nagahara},\ and\ \citenamefont {Okamoto}}]{INO05}%
  \BibitemOpen
  \bibfield  {author} {\bibinfo {author} {\bibfnamefont {Kohei}\ \bibnamefont
  {Imura}}, \bibinfo {author} {\bibfnamefont {Tetsuhiko}\ \bibnamefont
  {Nagahara}}, \ and\ \bibinfo {author} {\bibfnamefont {Hiromi}\ \bibnamefont
  {Okamoto}},\ }\bibfield  {title} {\enquote {\bibinfo {title} {Near-field
  two-photon-induced photoluminescence from single gold nanorods and imaging of
  plasmon modes},}\ }\href@noop {} {\bibfield  {journal} {\bibinfo  {journal}
  {J.\ Phys.\ Chem.\ B}\ }\textbf {\bibinfo {volume} {109}},\ \bibinfo {pages}
  {13214--13220} (\bibinfo {year} {2005})}\BibitemShut {NoStop}%
\bibitem [{\citenamefont {Bouhelier}\ \emph {et~al.}(2005)\citenamefont
  {Bouhelier}, \citenamefont {Bachelot}, \citenamefont {Lerondel},
  \citenamefont {Kostcheev}, \citenamefont {Royer},\ and\ \citenamefont
  {Wiederrecht}}]{BBL05}%
  \BibitemOpen
  \bibfield  {author} {\bibinfo {author} {\bibfnamefont {Alexandre}\
  \bibnamefont {Bouhelier}}, \bibinfo {author} {\bibfnamefont {Renaud}\
  \bibnamefont {Bachelot}}, \bibinfo {author} {\bibfnamefont {Gilles}\
  \bibnamefont {Lerondel}}, \bibinfo {author} {\bibfnamefont {Sergei}\
  \bibnamefont {Kostcheev}}, \bibinfo {author} {\bibfnamefont {Pascal}\
  \bibnamefont {Royer}}, \ and\ \bibinfo {author} {\bibfnamefont {G.~P.}\
  \bibnamefont {Wiederrecht}},\ }\bibfield  {title} {\enquote {\bibinfo {title}
  {Surface plasmon characteristics of tunable photoluminescence in single gold
  nanorods},}\ }\href@noop {} {\bibfield  {journal} {\bibinfo  {journal}
  {Phys.\ Rev.\ Lett.}\ }\textbf {\bibinfo {volume} {95}},\ \bibinfo {pages}
  {267405} (\bibinfo {year} {2005})}\BibitemShut {NoStop}%
\bibitem [{\citenamefont {Sheik-Bahae}\ and\ \citenamefont
  {Hasselbeck}(2000)}]{SH00}%
  \BibitemOpen
  \bibfield  {author} {\bibinfo {author} {\bibfnamefont {Mansoor}\ \bibnamefont
  {Sheik-Bahae}}\ and\ \bibinfo {author} {\bibfnamefont {Michael~P.}\
  \bibnamefont {Hasselbeck}},\ }\enquote {\bibinfo {title} {Third-order optical
  nonlinearities},}\ in\ \href@noop {} {\emph {\bibinfo {booktitle} {Optical
  Properties of Materials, Nonlinear Optics, Quantum Optics}}},\ \bibinfo
  {series} {Handbook of Optics}, Vol.~\bibinfo {volume} {IV}\ (\bibinfo
  {publisher} {McGraw-Hill Education},\ \bibinfo {address} {New York},\
  \bibinfo {year} {2000})\ Chap.~\bibinfo {chapter} {17}\BibitemShut {NoStop}%
\bibitem [{\citenamefont {Biagioni}\ \emph {et~al.}(2009)\citenamefont
  {Biagioni}, \citenamefont {Celebrano}, \citenamefont {Savoini}, \citenamefont
  {Grancini}, \citenamefont {Brida}, \citenamefont {M\'at\'efi-Tempfli},
  \citenamefont {M\'at\'efi-Tempfli}, \citenamefont {Du\`o}, \citenamefont
  {Hecht}, \citenamefont {Cerullo},\ and\ \citenamefont {Finazzi}}]{BCS09}%
  \BibitemOpen
  \bibfield  {author} {\bibinfo {author} {\bibfnamefont {P.}~\bibnamefont
  {Biagioni}}, \bibinfo {author} {\bibfnamefont {M.}~\bibnamefont {Celebrano}},
  \bibinfo {author} {\bibfnamefont {M.}~\bibnamefont {Savoini}}, \bibinfo
  {author} {\bibfnamefont {G.}~\bibnamefont {Grancini}}, \bibinfo {author}
  {\bibfnamefont {D.}~\bibnamefont {Brida}}, \bibinfo {author} {\bibfnamefont
  {S.}~\bibnamefont {M\'at\'efi-Tempfli}}, \bibinfo {author} {\bibfnamefont
  {M.}~\bibnamefont {M\'at\'efi-Tempfli}}, \bibinfo {author} {\bibfnamefont
  {L.}~\bibnamefont {Du\`o}}, \bibinfo {author} {\bibfnamefont
  {B.}~\bibnamefont {Hecht}}, \bibinfo {author} {\bibfnamefont
  {G.}~\bibnamefont {Cerullo}}, \ and\ \bibinfo {author} {\bibfnamefont
  {M.}~\bibnamefont {Finazzi}},\ }\bibfield  {title} {\enquote {\bibinfo
  {title} {Dependence of the two-photon photoluminescence yield of gold
  nanostructures on the laser pulse duration},}\ }\href@noop {} {\bibfield
  {journal} {\bibinfo  {journal} {Phys.\ Rev.\ B}\ }\textbf {\bibinfo {volume}
  {80}},\ \bibinfo {pages} {045411} (\bibinfo {year} {2009})}\BibitemShut
  {NoStop}%
\bibitem [{\citenamefont {Boroviks}\ \emph {et~al.}(2021)\citenamefont
  {Boroviks}, \citenamefont {Yezekyan}, \citenamefont {{Rodr{\'\i}guez
  Echarri}}, \citenamefont {{Garc{\'\i}a de Abajo}}, \citenamefont {Cox},
  \citenamefont {Bozhevolnyi}, \citenamefont {Mortensen},\ and\ \citenamefont
  {Wolff}}]{paper363}%
  \BibitemOpen
  \bibfield  {author} {\bibinfo {author} {\bibfnamefont {S.}~\bibnamefont
  {Boroviks}}, \bibinfo {author} {\bibfnamefont {T.}~\bibnamefont {Yezekyan}},
  \bibinfo {author} {\bibfnamefont {R.}~\bibnamefont {{Rodr{\'\i}guez
  Echarri}}}, \bibinfo {author} {\bibfnamefont {F.~J.}\ \bibnamefont
  {{Garc{\'\i}a de Abajo}}}, \bibinfo {author} {\bibfnamefont {J.~D.}\
  \bibnamefont {Cox}}, \bibinfo {author} {\bibfnamefont {S.~I..}\ \bibnamefont
  {Bozhevolnyi}}, \bibinfo {author} {\bibfnamefont {N.~A.}\ \bibnamefont
  {Mortensen}}, \ and\ \bibinfo {author} {\bibfnamefont {C.}~\bibnamefont
  {Wolff}},\ }\bibfield  {title} {\enquote {\bibinfo {title} {Anisotropic
  second-harmonic generation from monocrystalline gold flakes},}\ }\href@noop
  {} {\bibfield  {journal} {\bibinfo  {journal} {Opt.\ Lett.}\ }\textbf
  {\bibinfo {volume} {46}},\ \bibinfo {pages} {833--836} (\bibinfo {year}
  {2021})}\BibitemShut {NoStop}%
\bibitem [{\citenamefont {Muehlschlegel}\ \emph {et~al.}(2005)\citenamefont
  {Muehlschlegel}, \citenamefont {Eisler}, \citenamefont {Martin},
  \citenamefont {Hecht},\ and\ \citenamefont {Pohl}}]{MEM05}%
  \BibitemOpen
  \bibfield  {author} {\bibinfo {author} {\bibfnamefont {Peter}\ \bibnamefont
  {Muehlschlegel}}, \bibinfo {author} {\bibfnamefont {H.-J.}\ \bibnamefont
  {Eisler}}, \bibinfo {author} {\bibfnamefont {Olivier J.~F.}\ \bibnamefont
  {Martin}}, \bibinfo {author} {\bibfnamefont {Bert}\ \bibnamefont {Hecht}}, \
  and\ \bibinfo {author} {\bibfnamefont {D.~W.}\ \bibnamefont {Pohl}},\
  }\bibfield  {title} {\enquote {\bibinfo {title} {Resonant optical
  antennas},}\ }\href@noop {} {\bibfield  {journal} {\bibinfo  {journal}
  {Science}\ }\textbf {\bibinfo {volume} {308}},\ \bibinfo {pages} {1607--1609}
  (\bibinfo {year} {2005})}\BibitemShut {NoStop}%
\bibitem [{\citenamefont {Buhmann}\ \emph {et~al.}(2012)\citenamefont
  {Buhmann}, \citenamefont {Butcher},\ and\ \citenamefont {Scheel}}]{BBS12}%
  \BibitemOpen
  \bibfield  {author} {\bibinfo {author} {\bibfnamefont {S.~Y.}\ \bibnamefont
  {Buhmann}}, \bibinfo {author} {\bibfnamefont {D.~T.}\ \bibnamefont
  {Butcher}}, \ and\ \bibinfo {author} {\bibfnamefont {S.}~\bibnamefont
  {Scheel}},\ }\bibfield  {title} {\enquote {\bibinfo {title} {Macroscopic
  quantum electrodynamics in nonlocal and nonreciprocal media},}\ }\href@noop
  {} {\bibfield  {journal} {\bibinfo  {journal} {New\ J.\ Phys.}\ }\textbf
  {\bibinfo {volume} {14}},\ \bibinfo {pages} {083034} (\bibinfo {year}
  {2012})}\BibitemShut {NoStop}%
\bibitem [{\citenamefont {Biagioni}\ \emph {et~al.}(2012)\citenamefont
  {Biagioni}, \citenamefont {Brida}, \citenamefont {Huang}, \citenamefont
  {Kern}, \citenamefont {Du{\`o}}, \citenamefont {Hecht}, \citenamefont
  {Finazzi},\ and\ \citenamefont {Cerullo}}]{BBH12}%
  \BibitemOpen
  \bibfield  {author} {\bibinfo {author} {\bibfnamefont {Paolo}\ \bibnamefont
  {Biagioni}}, \bibinfo {author} {\bibfnamefont {Daniele}\ \bibnamefont
  {Brida}}, \bibinfo {author} {\bibfnamefont {Jer-Shing}\ \bibnamefont
  {Huang}}, \bibinfo {author} {\bibfnamefont {Johannes}\ \bibnamefont {Kern}},
  \bibinfo {author} {\bibfnamefont {Lamberto}\ \bibnamefont {Du{\`o}}},
  \bibinfo {author} {\bibfnamefont {Bert}\ \bibnamefont {Hecht}}, \bibinfo
  {author} {\bibfnamefont {Marco}\ \bibnamefont {Finazzi}}, \ and\ \bibinfo
  {author} {\bibfnamefont {Giulio}\ \bibnamefont {Cerullo}},\ }\bibfield
  {title} {\enquote {\bibinfo {title} {Dynamics of four-photon
  photoluminescence in gold nanoantennas},}\ }\href@noop {} {\bibfield
  {journal} {\bibinfo  {journal} {Nano\ Lett.}\ }\textbf {\bibinfo {volume}
  {12}},\ \bibinfo {pages} {2941--2947} (\bibinfo {year} {2012})}\BibitemShut
  {NoStop}%
\bibitem [{\citenamefont {Bauer}\ \emph {et~al.}(2015)\citenamefont {Bauer},
  \citenamefont {Marienfeld},\ and\ \citenamefont {Aeschlimann}}]{BMA15}%
  \BibitemOpen
  \bibfield  {author} {\bibinfo {author} {\bibfnamefont {M.}~\bibnamefont
  {Bauer}}, \bibinfo {author} {\bibfnamefont {A.}~\bibnamefont {Marienfeld}}, \
  and\ \bibinfo {author} {\bibfnamefont {M.}~\bibnamefont {Aeschlimann}},\
  }\bibfield  {title} {\enquote {\bibinfo {title} {Hot electron lifetimes in
  metals probed by time-resolved two-photon photoemission},}\ }\href {\doibase
  10.1016/j.progsurf.2015.05.001} {\bibfield  {journal} {\bibinfo  {journal}
  {Prog.\ Surf.\ Sci.}\ }\textbf {\bibinfo {volume} {90}},\ \bibinfo {pages}
  {319--376} (\bibinfo {year} {2015})}\BibitemShut {NoStop}%
\bibitem [{\citenamefont {Saavedra}\ \emph {et~al.}(2016)\citenamefont
  {Saavedra}, \citenamefont {Asenjo-Garcia},\ and\ \citenamefont {{Garc\'{\i}a
  de Abajo}}}]{paper280}%
  \BibitemOpen
  \bibfield  {author} {\bibinfo {author} {\bibfnamefont {J.~R.~M.}\
  \bibnamefont {Saavedra}}, \bibinfo {author} {\bibfnamefont {A.}~\bibnamefont
  {Asenjo-Garcia}}, \ and\ \bibinfo {author} {\bibfnamefont {F.~J.}\
  \bibnamefont {{Garc\'{\i}a de Abajo}}},\ }\bibfield  {title} {\enquote
  {\bibinfo {title} {Hot-electron dynamics and thermalization in small metallic
  nanoparticles},}\ }\href@noop {} {\bibfield  {journal} {\bibinfo  {journal}
  {ACS\ Photonics}\ }\textbf {\bibinfo {volume} {3}},\ \bibinfo {pages}
  {1637--1646} (\bibinfo {year} {2016})}\BibitemShut {NoStop}%
\bibitem [{\citenamefont {Dubi}\ and\ \citenamefont {Sivan}(2019)}]{DS19}%
  \BibitemOpen
  \bibfield  {author} {\bibinfo {author} {\bibfnamefont {Yonatan}\ \bibnamefont
  {Dubi}}\ and\ \bibinfo {author} {\bibfnamefont {Yonatan}\ \bibnamefont
  {Sivan}},\ }\bibfield  {title} {\enquote {\bibinfo {title} {“hot”
  electrons in metallic nanostructures—non-thermal carriers or heating?}}\
  }\href@noop {} {\bibfield  {journal} {\bibinfo  {journal} {Light\ Sci.\
  Appl.}\ }\textbf {\bibinfo {volume} {8}},\ \bibinfo {pages} {89} (\bibinfo
  {year} {2019})}\BibitemShut {NoStop}%
\bibitem [{\citenamefont {Haug}\ \emph {et~al.}(2015)\citenamefont {Haug},
  \citenamefont {Klemm}, \citenamefont {Bange},\ and\ \citenamefont
  {Lupton}}]{HKB15}%
  \BibitemOpen
  \bibfield  {author} {\bibinfo {author} {\bibfnamefont {Tobias}\ \bibnamefont
  {Haug}}, \bibinfo {author} {\bibfnamefont {Philippe}\ \bibnamefont {Klemm}},
  \bibinfo {author} {\bibfnamefont {Sebastian}\ \bibnamefont {Bange}}, \ and\
  \bibinfo {author} {\bibfnamefont {John~M}\ \bibnamefont {Lupton}},\
  }\bibfield  {title} {\enquote {\bibinfo {title} {Hot-electron intraband
  luminescence from single hot spots in noble-metal nanoparticle films},}\
  }\href@noop {} {\bibfield  {journal} {\bibinfo  {journal} {Phys.\ Rev.\
  Lett.}\ }\textbf {\bibinfo {volume} {115}},\ \bibinfo {pages} {067403}
  (\bibinfo {year} {2015})}\BibitemShut {NoStop}%
\bibitem [{\citenamefont {Fowler}(1931)}]{F1931}%
  \BibitemOpen
  \bibfield  {author} {\bibinfo {author} {\bibfnamefont {Ralph~H.}\
  \bibnamefont {Fowler}},\ }\bibfield  {title} {\enquote {\bibinfo {title} {The
  analysis of photoelectric sensitivity curves for clean metals at various
  temperatures},}\ }\href@noop {} {\bibfield  {journal} {\bibinfo  {journal}
  {Phys.\ Rev.}\ }\textbf {\bibinfo {volume} {38}},\ \bibinfo {pages} {45}
  (\bibinfo {year} {1931})}\BibitemShut {NoStop}%
\bibitem [{\citenamefont {DuBridge}(1932)}]{D1932}%
  \BibitemOpen
  \bibfield  {author} {\bibinfo {author} {\bibfnamefont {Lee~A.}\ \bibnamefont
  {DuBridge}},\ }\bibfield  {title} {\enquote {\bibinfo {title} {A further
  experimental test of fowler's theory of photoelectric emission},}\
  }\href@noop {} {\bibfield  {journal} {\bibinfo  {journal} {Phys.\ Rev.}\
  }\textbf {\bibinfo {volume} {39}},\ \bibinfo {pages} {108} (\bibinfo {year}
  {1932})}\BibitemShut {NoStop}%
\bibitem [{\citenamefont {DuBridge}(1933)}]{D1933}%
  \BibitemOpen
  \bibfield  {author} {\bibinfo {author} {\bibfnamefont {Lee~A.}\ \bibnamefont
  {DuBridge}},\ }\bibfield  {title} {\enquote {\bibinfo {title} {Theory of the
  energy distribution of photoelectrons},}\ }\href@noop {} {\bibfield
  {journal} {\bibinfo  {journal} {Phys.\ Rev.}\ }\textbf {\bibinfo {volume}
  {43}},\ \bibinfo {pages} {727} (\bibinfo {year} {1933})}\BibitemShut
  {NoStop}%
\bibitem [{\citenamefont {Zhou}\ and\ \citenamefont {Zhang}(2020)}]{ZZ20}%
  \BibitemOpen
  \bibfield  {author} {\bibinfo {author} {\bibfnamefont {Yang}\ \bibnamefont
  {Zhou}}\ and\ \bibinfo {author} {\bibfnamefont {Peng}\ \bibnamefont
  {Zhang}},\ }\bibfield  {title} {\enquote {\bibinfo {title} {A quantum model
  for photoemission from metal surfaces and its comparison with the three-step
  model and fowler--dubridge model},}\ }\href@noop {} {\bibfield  {journal}
  {\bibinfo  {journal} {J.\ Appl.\ Phys.}\ }\textbf {\bibinfo {volume} {127}},\
  \bibinfo {pages} {164903} (\bibinfo {year} {2020})}\BibitemShut {NoStop}%
\bibitem [{\citenamefont {Mejard}\ \emph {et~al.}(2016)\citenamefont {Mejard},
  \citenamefont {Verdy}, \citenamefont {Petit}, \citenamefont {Bouhelier},
  \citenamefont {Cluzel},\ and\ \citenamefont {Demichel}}]{MVP16}%
  \BibitemOpen
  \bibfield  {author} {\bibinfo {author} {\bibfnamefont {Regis}\ \bibnamefont
  {Mejard}}, \bibinfo {author} {\bibfnamefont {Anthonin}\ \bibnamefont
  {Verdy}}, \bibinfo {author} {\bibfnamefont {Marlene}\ \bibnamefont {Petit}},
  \bibinfo {author} {\bibfnamefont {Alexandre}\ \bibnamefont {Bouhelier}},
  \bibinfo {author} {\bibfnamefont {Benoit}\ \bibnamefont {Cluzel}}, \ and\
  \bibinfo {author} {\bibfnamefont {Olivier}\ \bibnamefont {Demichel}},\
  }\bibfield  {title} {\enquote {\bibinfo {title} {Energy-resolved hot-carrier
  relaxation dynamics in monocrystalline plasmonic nanoantennas},}\ }\href@noop
  {} {\bibfield  {journal} {\bibinfo  {journal} {ACS\ Photonics}\ }\textbf
  {\bibinfo {volume} {3}},\ \bibinfo {pages} {1482--1488} (\bibinfo {year}
  {2016})}\BibitemShut {NoStop}%
\bibitem [{\citenamefont {Roloff}\ \emph {et~al.}(2017)\citenamefont {Roloff},
  \citenamefont {Klemm}, \citenamefont {Gronwald}, \citenamefont {Huber},
  \citenamefont {Lupton},\ and\ \citenamefont {Bange}}]{RKG17}%
  \BibitemOpen
  \bibfield  {author} {\bibinfo {author} {\bibfnamefont {Lukas}\ \bibnamefont
  {Roloff}}, \bibinfo {author} {\bibfnamefont {Philippe}\ \bibnamefont
  {Klemm}}, \bibinfo {author} {\bibfnamefont {Imke}\ \bibnamefont {Gronwald}},
  \bibinfo {author} {\bibfnamefont {Rupert}\ \bibnamefont {Huber}}, \bibinfo
  {author} {\bibfnamefont {John~M}\ \bibnamefont {Lupton}}, \ and\ \bibinfo
  {author} {\bibfnamefont {Sebastian}\ \bibnamefont {Bange}},\ }\bibfield
  {title} {\enquote {\bibinfo {title} {Light emission from gold nanoparticles
  under ultrafast near-infrared excitation: Thermal radiation, inelastic light
  scattering, or multiphoton luminescence?}}\ }\href@noop {} {\bibfield
  {journal} {\bibinfo  {journal} {Nano\ Lett.}\ }\textbf {\bibinfo {volume}
  {17}},\ \bibinfo {pages} {7914--7919} (\bibinfo {year} {2017})}\BibitemShut
  {NoStop}%
\bibitem [{\citenamefont {Jollans}\ \emph {et~al.}(2020)\citenamefont
  {Jollans}, \citenamefont {Caldarola}, \citenamefont {Sivan},\ and\
  \citenamefont {Orrit}}]{JCS20}%
  \BibitemOpen
  \bibfield  {author} {\bibinfo {author} {\bibfnamefont {Thomas}\ \bibnamefont
  {Jollans}}, \bibinfo {author} {\bibfnamefont {Mart{\'\i}n}\ \bibnamefont
  {Caldarola}}, \bibinfo {author} {\bibfnamefont {Yonatan}\ \bibnamefont
  {Sivan}}, \ and\ \bibinfo {author} {\bibfnamefont {Michel}\ \bibnamefont
  {Orrit}},\ }\bibfield  {title} {\enquote {\bibinfo {title} {Effective
  electron temperature measurement using time-resolved anti-stokes
  photoluminescence},}\ }\href@noop {} {\bibfield  {journal} {\bibinfo
  {journal} {J.\ Phys.\ Chem.\ A}\ }\textbf {\bibinfo {volume} {124}},\
  \bibinfo {pages} {6968--6976} (\bibinfo {year} {2020})}\BibitemShut {NoStop}%
\bibitem [{\citenamefont {{Garc\'{\i}a de Abajo}}(2014)}]{paper235}%
  \BibitemOpen
  \bibfield  {author} {\bibinfo {author} {\bibfnamefont {F.~J.}\ \bibnamefont
  {{Garc\'{\i}a de Abajo}}},\ }\bibfield  {title} {\enquote {\bibinfo {title}
  {Graphene plasmonics: challenges and opportunities},}\ }\href {\doibase
  10.1021/ph400147y} {\bibfield  {journal} {\bibinfo  {journal} {ACS\
  Photonics}\ }\textbf {\bibinfo {volume} {1}},\ \bibinfo {pages} {135--152}
  (\bibinfo {year} {2014})}\BibitemShut {NoStop}%
\bibitem [{\citenamefont {{Rodr\'{\i}guez Echarri}}\ \emph
  {et~al.}(2021)\citenamefont {{Rodr\'{\i}guez Echarri}}, \citenamefont {Cox},
  \citenamefont {Iyikanat},\ and\ \citenamefont {{Garc\'{\i}a de
  Abajo}}}]{paper382}%
  \BibitemOpen
  \bibfield  {author} {\bibinfo {author} {\bibfnamefont {A.}~\bibnamefont
  {{Rodr\'{\i}guez Echarri}}}, \bibinfo {author} {\bibfnamefont {J.~D.}\
  \bibnamefont {Cox}}, \bibinfo {author} {\bibfnamefont {F.}~\bibnamefont
  {Iyikanat}}, \ and\ \bibinfo {author} {\bibfnamefont {F.~J.}\ \bibnamefont
  {{Garc\'{\i}a de Abajo}}},\ }\bibfield  {title} {\enquote {\bibinfo {title}
  {Nonlinear plasmonic response in atomically thin metal films},}\ }\href@noop
  {} {\bibfield  {journal} {\bibinfo  {journal} {Nanophotonics}\ }\textbf
  {\bibinfo {volume} {10}},\ \bibinfo {pages} {4149--4159} (\bibinfo {year}
  {2021})}\BibitemShut {NoStop}%
\bibitem [{\citenamefont {Boltasseva}\ and\ \citenamefont
  {Shalaev}(2019)}]{BS19}%
  \BibitemOpen
  \bibfield  {author} {\bibinfo {author} {\bibfnamefont {Alexandra}\
  \bibnamefont {Boltasseva}}\ and\ \bibinfo {author} {\bibfnamefont
  {Vladimir~M.}\ \bibnamefont {Shalaev}},\ }\bibfield  {title} {\enquote
  {\bibinfo {title} {Transdimensional photonics},}\ }\href@noop {} {\bibfield
  {journal} {\bibinfo  {journal} {ACS\ Photonics}\ }\textbf {\bibinfo {volume}
  {6}},\ \bibinfo {pages} {1--3} (\bibinfo {year} {2019})}\BibitemShut
  {NoStop}%
\bibitem [{\citenamefont {Lee}\ \emph {et~al.}(2014)\citenamefont {Lee},
  \citenamefont {Tymchenko}, \citenamefont {Argyropoulos}, \citenamefont
  {Chen}, \citenamefont {Lu}, \citenamefont {Demmerle}, \citenamefont {Boehm},
  \citenamefont {Amann}, \citenamefont {Al\'u},\ and\ \citenamefont
  {Belkin}}]{LTA14}%
  \BibitemOpen
  \bibfield  {author} {\bibinfo {author} {\bibfnamefont {Jongwon}\ \bibnamefont
  {Lee}}, \bibinfo {author} {\bibfnamefont {Mykhailo}\ \bibnamefont
  {Tymchenko}}, \bibinfo {author} {\bibfnamefont {Christos}\ \bibnamefont
  {Argyropoulos}}, \bibinfo {author} {\bibfnamefont {Pai-Yen}\ \bibnamefont
  {Chen}}, \bibinfo {author} {\bibfnamefont {Feng}\ \bibnamefont {Lu}},
  \bibinfo {author} {\bibfnamefont {Frederic}\ \bibnamefont {Demmerle}},
  \bibinfo {author} {\bibfnamefont {Gerhard}\ \bibnamefont {Boehm}}, \bibinfo
  {author} {\bibfnamefont {Markus-Christian}\ \bibnamefont {Amann}}, \bibinfo
  {author} {\bibfnamefont {Andrea}\ \bibnamefont {Al\'u}}, \ and\ \bibinfo
  {author} {\bibfnamefont {Mikhail~A.}\ \bibnamefont {Belkin}},\ }\bibfield
  {title} {\enquote {\bibinfo {title} {Giant nonlinear response from plasmonic
  metasurfaces coupled to intersubband transitions},}\ }\href@noop {}
  {\bibfield  {journal} {\bibinfo  {journal} {Nature}\ }\textbf {\bibinfo
  {volume} {511}},\ \bibinfo {pages} {65--69} (\bibinfo {year}
  {2014})}\BibitemShut {NoStop}%
\bibitem [{\citenamefont {Kubo}\ \emph {et~al.}(2005)\citenamefont {Kubo},
  \citenamefont {Onda}, \citenamefont {Petek}, \citenamefont {Sun},
  \citenamefont {Jung},\ and\ \citenamefont {Kim}}]{KOP05}%
  \BibitemOpen
  \bibfield  {author} {\bibinfo {author} {\bibfnamefont {A.}~\bibnamefont
  {Kubo}}, \bibinfo {author} {\bibfnamefont {K.}~\bibnamefont {Onda}}, \bibinfo
  {author} {\bibfnamefont {H.}~\bibnamefont {Petek}}, \bibinfo {author}
  {\bibfnamefont {Z.}~\bibnamefont {Sun}}, \bibinfo {author} {\bibfnamefont
  {Y.~S.}\ \bibnamefont {Jung}}, \ and\ \bibinfo {author} {\bibfnamefont
  {H.~K.}\ \bibnamefont {Kim}},\ }\bibfield  {title} {\enquote {\bibinfo
  {title} {Femtosecond imaging of surface plasmon dynamics in a nanostructured
  silver film},}\ }\href@noop {} {\bibfield  {journal} {\bibinfo  {journal}
  {Nano\ Lett.}\ }\textbf {\bibinfo {volume} {5}},\ \bibinfo {pages}
  {1123--1127} (\bibinfo {year} {2005})}\BibitemShut {NoStop}%
\bibitem [{\citenamefont {Block}\ \emph {et~al.}(2019)\citenamefont {Block},
  \citenamefont {Liebel}, \citenamefont {Yu}, \citenamefont {Spector},
  \citenamefont {Sivan}, \citenamefont {{Garc\'{\i}a de Abajo}},\ and\
  \citenamefont {{van Hulst}}}]{paper330}%
  \BibitemOpen
  \bibfield  {author} {\bibinfo {author} {\bibfnamefont {A.}~\bibnamefont
  {Block}}, \bibinfo {author} {\bibfnamefont {M.}~\bibnamefont {Liebel}},
  \bibinfo {author} {\bibfnamefont {R.}~\bibnamefont {Yu}}, \bibinfo {author}
  {\bibfnamefont {R.}~\bibnamefont {Spector}}, \bibinfo {author} {\bibfnamefont
  {Y.}~\bibnamefont {Sivan}}, \bibinfo {author} {\bibfnamefont {F.~J.}\
  \bibnamefont {{Garc\'{\i}a de Abajo}}}, \ and\ \bibinfo {author}
  {\bibfnamefont {N.~F.}\ \bibnamefont {{van Hulst}}},\ }\bibfield  {title}
  {\enquote {\bibinfo {title} {Tracking ultrafast hot-electron diffusion in
  space and time by ultrafast thermomodulation microscopy},}\ }\href@noop {}
  {\bibfield  {journal} {\bibinfo  {journal} {Sci.\ Adv.}\ }\textbf {\bibinfo
  {volume} {5}},\ \bibinfo {pages} {eaav8965} (\bibinfo {year}
  {2019})}\BibitemShut {NoStop}%
\bibitem [{\citenamefont {Chen}\ \emph {et~al.}(2010)\citenamefont {Chen},
  \citenamefont {Xu}, \citenamefont {Jiang}, \citenamefont {Sui}, \citenamefont
  {Ding}, \citenamefont {Liu},\ and\ \citenamefont {Jin}}]{CXJ10}%
  \BibitemOpen
  \bibfield  {author} {\bibinfo {author} {\bibfnamefont {AM}~\bibnamefont
  {Chen}}, \bibinfo {author} {\bibfnamefont {HF}~\bibnamefont {Xu}}, \bibinfo
  {author} {\bibfnamefont {YF}~\bibnamefont {Jiang}}, \bibinfo {author}
  {\bibfnamefont {LZ}~\bibnamefont {Sui}}, \bibinfo {author} {\bibfnamefont
  {DJ}~\bibnamefont {Ding}}, \bibinfo {author} {\bibfnamefont {H}~\bibnamefont
  {Liu}}, \ and\ \bibinfo {author} {\bibfnamefont {MX}~\bibnamefont {Jin}},\
  }\bibfield  {title} {\enquote {\bibinfo {title} {Modeling of femtosecond
  laser damage threshold on the two-layer metal films},}\ }\href@noop {}
  {\bibfield  {journal} {\bibinfo  {journal} {Appl.\ Surf.\ Sci.}\ }\textbf
  {\bibinfo {volume} {257}},\ \bibinfo {pages} {1678--1683} (\bibinfo {year}
  {2010})}\BibitemShut {NoStop}%
\bibitem [{\citenamefont {Zhang}\ \emph {et~al.}(2015)\citenamefont {Zhang},
  \citenamefont {Chen}, \citenamefont {Hu},\ and\ \citenamefont
  {Chen}}]{ZCH15}%
  \BibitemOpen
  \bibfield  {author} {\bibinfo {author} {\bibfnamefont {Jinping}\ \bibnamefont
  {Zhang}}, \bibinfo {author} {\bibfnamefont {Yuping}\ \bibnamefont {Chen}},
  \bibinfo {author} {\bibfnamefont {Mengning}\ \bibnamefont {Hu}}, \ and\
  \bibinfo {author} {\bibfnamefont {Xianfeng}\ \bibnamefont {Chen}},\
  }\bibfield  {title} {\enquote {\bibinfo {title} {An improved
  three-dimensional two-temperature model for multi-pulse femtosecond laser
  ablation of aluminum},}\ }\href@noop {} {\bibfield  {journal} {\bibinfo
  {journal} {J.\ Appl.\ Phys.}\ }\textbf {\bibinfo {volume} {117}},\ \bibinfo
  {pages} {063104} (\bibinfo {year} {2015})}\BibitemShut {NoStop}%
\bibitem [{\citenamefont {Rethfeld}\ \emph {et~al.}(2017)\citenamefont
  {Rethfeld}, \citenamefont {Ivanov}, \citenamefont {Garcia},\ and\
  \citenamefont {Anisimov}}]{RIG17}%
  \BibitemOpen
  \bibfield  {author} {\bibinfo {author} {\bibfnamefont {Baerbel}\ \bibnamefont
  {Rethfeld}}, \bibinfo {author} {\bibfnamefont {Dmitriy~S.}\ \bibnamefont
  {Ivanov}}, \bibinfo {author} {\bibfnamefont {Martin~E.}\ \bibnamefont
  {Garcia}}, \ and\ \bibinfo {author} {\bibfnamefont {Sergei~I.}\ \bibnamefont
  {Anisimov}},\ }\bibfield  {title} {\enquote {\bibinfo {title} {Modelling
  ultrafast laser ablation},}\ }\href@noop {} {\bibfield  {journal} {\bibinfo
  {journal} {J.\ Phys.\ D}\ }\textbf {\bibinfo {volume} {50}},\ \bibinfo
  {pages} {193001} (\bibinfo {year} {2017})}\BibitemShut {NoStop}%
\bibitem [{\citenamefont {Wang}\ \emph {et~al.}(1994)\citenamefont {Wang},
  \citenamefont {Riffe}, \citenamefont {Lee},\ and\ \citenamefont
  {Downer}}]{WRL94}%
  \BibitemOpen
  \bibfield  {author} {\bibinfo {author} {\bibfnamefont {X.~Y.}\ \bibnamefont
  {Wang}}, \bibinfo {author} {\bibfnamefont {D.~M.}\ \bibnamefont {Riffe}},
  \bibinfo {author} {\bibfnamefont {Y.-S.}\ \bibnamefont {Lee}}, \ and\
  \bibinfo {author} {\bibfnamefont {M.~C.}\ \bibnamefont {Downer}},\ }\bibfield
   {title} {\enquote {\bibinfo {title} {Time-resolved electron-temperature
  measurement in a highly excited gold target using femtosecond thermionic
  emission},}\ }\href {\doibase 10.1103/PhysRevB.50.8016} {\bibfield  {journal}
  {\bibinfo  {journal} {Phys.\ Rev.\ B}\ }\textbf {\bibinfo {volume} {50}},\
  \bibinfo {pages} {8016--8019} (\bibinfo {year} {1994})}\BibitemShut {NoStop}%
\bibitem [{\citenamefont {Sakat}\ \emph {et~al.}(2016)\citenamefont {Sakat},
  \citenamefont {Bargigia}, \citenamefont {Celebrano}, \citenamefont {Cattoni},
  \citenamefont {Collin}, \citenamefont {Brida}, \citenamefont {Finazzi},
  \citenamefont {D'~Andrea},\ and\ \citenamefont {Biagioni}}]{SBC16}%
  \BibitemOpen
  \bibfield  {author} {\bibinfo {author} {\bibfnamefont {Emilie}\ \bibnamefont
  {Sakat}}, \bibinfo {author} {\bibfnamefont {Ilaria}\ \bibnamefont
  {Bargigia}}, \bibinfo {author} {\bibfnamefont {Michele}\ \bibnamefont
  {Celebrano}}, \bibinfo {author} {\bibfnamefont {Andrea}\ \bibnamefont
  {Cattoni}}, \bibinfo {author} {\bibfnamefont {St{\'e}phane}\ \bibnamefont
  {Collin}}, \bibinfo {author} {\bibfnamefont {Daniele}\ \bibnamefont {Brida}},
  \bibinfo {author} {\bibfnamefont {Marco}\ \bibnamefont {Finazzi}}, \bibinfo
  {author} {\bibfnamefont {Cosimo}\ \bibnamefont {D'~Andrea}}, \ and\ \bibinfo
  {author} {\bibfnamefont {Paolo}\ \bibnamefont {Biagioni}},\ }\bibfield
  {title} {\enquote {\bibinfo {title} {Time-resolved photoluminescence in gold
  nanoantennas},}\ }\href@noop {} {\bibfield  {journal} {\bibinfo  {journal}
  {ACS\ Photonics}\ }\textbf {\bibinfo {volume} {3}},\ \bibinfo {pages}
  {1489--1493} (\bibinfo {year} {2016})}\BibitemShut {NoStop}%
\bibitem [{\citenamefont {Ashcroft}\ and\ \citenamefont
  {Mermin}(1976)}]{AM1976}%
  \BibitemOpen
  \bibfield  {author} {\bibinfo {author} {\bibfnamefont {N.~W.}\ \bibnamefont
  {Ashcroft}}\ and\ \bibinfo {author} {\bibfnamefont {N.~D.}\ \bibnamefont
  {Mermin}},\ }\href@noop {} {\emph {\bibinfo {title} {Solid State Physics}}}\
  (\bibinfo  {publisher} {Harcourt College Publishers},\ \bibinfo {address}
  {Philadelphia},\ \bibinfo {year} {1976})\BibitemShut {NoStop}%
\bibitem [{\citenamefont {Bonn}\ \emph {et~al.}(2000)\citenamefont {Bonn},
  \citenamefont {Denzler}, \citenamefont {Funk}, \citenamefont {Wolf},
  \citenamefont {Wellershoff},\ and\ \citenamefont {Hohlfeld}}]{BDF00}%
  \BibitemOpen
  \bibfield  {author} {\bibinfo {author} {\bibfnamefont {Mischa}\ \bibnamefont
  {Bonn}}, \bibinfo {author} {\bibfnamefont {Daniel~N.}\ \bibnamefont
  {Denzler}}, \bibinfo {author} {\bibfnamefont {Stephan}\ \bibnamefont {Funk}},
  \bibinfo {author} {\bibfnamefont {Martin}\ \bibnamefont {Wolf}}, \bibinfo
  {author} {\bibfnamefont {S.-Svante}\ \bibnamefont {Wellershoff}}, \ and\
  \bibinfo {author} {\bibfnamefont {Julius}\ \bibnamefont {Hohlfeld}},\
  }\bibfield  {title} {\enquote {\bibinfo {title} {Ultrafast electron dynamics
  at metal surfaces: Competition between electron-phonon coupling and
  hot-electron transport},}\ }\href@noop {} {\bibfield  {journal} {\bibinfo
  {journal} {Phys.\ Rev.\ B}\ }\textbf {\bibinfo {volume} {61}},\ \bibinfo
  {pages} {1101} (\bibinfo {year} {2000})}\BibitemShut {NoStop}%
\bibitem [{\citenamefont {Lee}\ \emph {et~al.}(2011)\citenamefont {Lee},
  \citenamefont {Kang},\ and\ \citenamefont {Lee}}]{LKL11}%
  \BibitemOpen
  \bibfield  {author} {\bibinfo {author} {\bibfnamefont {Jae~Bin}\ \bibnamefont
  {Lee}}, \bibinfo {author} {\bibfnamefont {Kwangu}\ \bibnamefont {Kang}}, \
  and\ \bibinfo {author} {\bibfnamefont {Seong~Hyuk}\ \bibnamefont {Lee}},\
  }\bibfield  {title} {\enquote {\bibinfo {title} {Comparison of theoretical
  models of electron-phonon coupling in thin gold films irradiated by
  femtosecond pulse lasers},}\ }\href@noop {} {\bibfield  {journal} {\bibinfo
  {journal} {Mater.\ Trans.}\ }\textbf {\bibinfo {volume} {52}},\ \bibinfo
  {pages} {547--553} (\bibinfo {year} {2011})}\BibitemShut {NoStop}%
\bibitem [{\citenamefont {Johnson}\ and\ \citenamefont
  {Christy}(1972)}]{JC1972}%
  \BibitemOpen
  \bibfield  {author} {\bibinfo {author} {\bibfnamefont {P.~B.}\ \bibnamefont
  {Johnson}}\ and\ \bibinfo {author} {\bibfnamefont {R.~W.}\ \bibnamefont
  {Christy}},\ }\bibfield  {title} {\enquote {\bibinfo {title} {Optical
  constants of the noble metals},}\ }\href {\doibase 10.1103/PhysRevB.6.4370}
  {\bibfield  {journal} {\bibinfo  {journal} {Phys.\ Rev.\ B}\ }\textbf
  {\bibinfo {volume} {6}},\ \bibinfo {pages} {4370--4379} (\bibinfo {year}
  {1972})}\BibitemShut {NoStop}%
\bibitem [{\citenamefont {Lawrence}(1976)}]{L1976}%
  \BibitemOpen
  \bibfield  {author} {\bibinfo {author} {\bibfnamefont {W.~E.}\ \bibnamefont
  {Lawrence}},\ }\bibfield  {title} {\enquote {\bibinfo {title}
  {Electron-electron scattering in the low-temperature resistivity of the noble
  metals},}\ }\href@noop {} {\bibfield  {journal} {\bibinfo  {journal} {Phys.\
  Rev.\ B}\ }\textbf {\bibinfo {volume} {13}},\ \bibinfo {pages} {5316--5319}
  (\bibinfo {year} {1976})}\BibitemShut {NoStop}%
\bibitem [{\citenamefont {Beach}\ and\ \citenamefont {Christy}(1977)}]{BC1977}%
  \BibitemOpen
  \bibfield  {author} {\bibinfo {author} {\bibfnamefont {R.~T.}\ \bibnamefont
  {Beach}}\ and\ \bibinfo {author} {\bibfnamefont {R.~W.}\ \bibnamefont
  {Christy}},\ }\bibfield  {title} {\enquote {\bibinfo {title}
  {Electron-electron scattering in the intraband optical conductivity of {Cu},
  {Ag}, and {Au}},}\ }\href@noop {} {\bibfield  {journal} {\bibinfo  {journal}
  {Phys.\ Rev.\ B}\ }\textbf {\bibinfo {volume} {16}},\ \bibinfo {pages}
  {5277--5284} (\bibinfo {year} {1977})}\BibitemShut {NoStop}%
\bibitem [{\citenamefont {{O'Reilly}}(2017)}]{OE17}%
  \BibitemOpen
  \bibfield  {author} {\bibinfo {author} {\bibfnamefont {E.~P.}\ \bibnamefont
  {{O'Reilly}}},\ }\href@noop {} {\emph {\bibinfo {title} {Quantum theory of
  solids}}}\ (\bibinfo  {publisher} {Taylor \& Francis},\ \bibinfo {address}
  {New York},\ \bibinfo {year} {2017})\BibitemShut {NoStop}%
\bibitem [{\citenamefont {Boyd}(2008)}]{B08_3}%
  \BibitemOpen
  \bibfield  {author} {\bibinfo {author} {\bibfnamefont {Robert~W.}\
  \bibnamefont {Boyd}},\ }\href@noop {} {\emph {\bibinfo {title} {Nonlinear
  Optics}}},\ \bibinfo {edition} {3rd}\ ed.\ (\bibinfo  {publisher} {Academic
  Press},\ \bibinfo {address} {Amsterdam},\ \bibinfo {year} {2008})\BibitemShut
  {NoStop}%
\bibitem [{\citenamefont {Saleh}\ and\ \citenamefont {Teich}(2019)}]{ST19}%
  \BibitemOpen
  \bibfield  {author} {\bibinfo {author} {\bibfnamefont {Bahaa~EA}\
  \bibnamefont {Saleh}}\ and\ \bibinfo {author} {\bibfnamefont {Malvin~Carl}\
  \bibnamefont {Teich}},\ }\href@noop {} {\emph {\bibinfo {title} {Fundamentals
  of photonics}}}\ (\bibinfo  {publisher} {John Wiley \& Sons},\ \bibinfo
  {address} {New York},\ \bibinfo {year} {2019})\BibitemShut {NoStop}%
\bibitem [{\citenamefont {Giannozzi}\ \emph {et~al.}(2009)\citenamefont
  {Giannozzi}, \citenamefont {Baroni}, \citenamefont {Bonini}, \citenamefont
  {Calandra}, \citenamefont {Car}, \citenamefont {Cavazzoni}, \citenamefont
  {Ceresoli}, \citenamefont {Chiarotti}, \citenamefont {Cococcioni},
  \citenamefont {Dabo}, \citenamefont {Corso}, \citenamefont {Gironcoli},
  \citenamefont {Fabris}, \citenamefont {Fratesi}, \citenamefont {Gebauer},
  \citenamefont {Gerstmann}, \citenamefont {Gougoussis}, \citenamefont
  {Kokalj}, \citenamefont {Lazzeri}, \citenamefont {Martin-Samos},
  \citenamefont {Marzari}, \citenamefont {Mauri}, \citenamefont {Mazzarello},
  \citenamefont {Paolini}, \citenamefont {Pasquarello}, \citenamefont
  {Paulatto}, \citenamefont {Sbraccia}, \citenamefont {Scandolo}, \citenamefont
  {Sclauzero}, \citenamefont {Seitsonen}, \citenamefont {Smogunov},
  \citenamefont {Umari},\ and\ \citenamefont {Wentzcovitch}}]{GBB09}%
  \BibitemOpen
  \bibfield  {author} {\bibinfo {author} {\bibfnamefont {Paolo}\ \bibnamefont
  {Giannozzi}}, \bibinfo {author} {\bibfnamefont {Stefano}\ \bibnamefont
  {Baroni}}, \bibinfo {author} {\bibfnamefont {Nicola}\ \bibnamefont {Bonini}},
  \bibinfo {author} {\bibfnamefont {Matteo}\ \bibnamefont {Calandra}}, \bibinfo
  {author} {\bibfnamefont {Roberto}\ \bibnamefont {Car}}, \bibinfo {author}
  {\bibfnamefont {Carlo}\ \bibnamefont {Cavazzoni}}, \bibinfo {author}
  {\bibfnamefont {Davide}\ \bibnamefont {Ceresoli}}, \bibinfo {author}
  {\bibfnamefont {Guido~L.}\ \bibnamefont {Chiarotti}}, \bibinfo {author}
  {\bibfnamefont {Matteo}\ \bibnamefont {Cococcioni}}, \bibinfo {author}
  {\bibfnamefont {Ismaila}\ \bibnamefont {Dabo}}, \bibinfo {author}
  {\bibfnamefont {Andrea~Dal}\ \bibnamefont {Corso}}, \bibinfo {author}
  {\bibfnamefont {Stefano~de}\ \bibnamefont {Gironcoli}}, \bibinfo {author}
  {\bibfnamefont {Stefano}\ \bibnamefont {Fabris}}, \bibinfo {author}
  {\bibfnamefont {Guido}\ \bibnamefont {Fratesi}}, \bibinfo {author}
  {\bibfnamefont {Ralph}\ \bibnamefont {Gebauer}}, \bibinfo {author}
  {\bibfnamefont {Uwe}\ \bibnamefont {Gerstmann}}, \bibinfo {author}
  {\bibfnamefont {Christos}\ \bibnamefont {Gougoussis}}, \bibinfo {author}
  {\bibfnamefont {Anton}\ \bibnamefont {Kokalj}}, \bibinfo {author}
  {\bibfnamefont {Michele}\ \bibnamefont {Lazzeri}}, \bibinfo {author}
  {\bibfnamefont {Layla}\ \bibnamefont {Martin-Samos}}, \bibinfo {author}
  {\bibfnamefont {Nicola}\ \bibnamefont {Marzari}}, \bibinfo {author}
  {\bibfnamefont {Francesco}\ \bibnamefont {Mauri}}, \bibinfo {author}
  {\bibfnamefont {Riccardo}\ \bibnamefont {Mazzarello}}, \bibinfo {author}
  {\bibfnamefont {Stefano}\ \bibnamefont {Paolini}}, \bibinfo {author}
  {\bibfnamefont {Alfredo}\ \bibnamefont {Pasquarello}}, \bibinfo {author}
  {\bibfnamefont {Lorenzo}\ \bibnamefont {Paulatto}}, \bibinfo {author}
  {\bibfnamefont {Carlo}\ \bibnamefont {Sbraccia}}, \bibinfo {author}
  {\bibfnamefont {Sandro}\ \bibnamefont {Scandolo}}, \bibinfo {author}
  {\bibfnamefont {Gabriele}\ \bibnamefont {Sclauzero}}, \bibinfo {author}
  {\bibfnamefont {Ari~P.}\ \bibnamefont {Seitsonen}}, \bibinfo {author}
  {\bibfnamefont {Alexander}\ \bibnamefont {Smogunov}}, \bibinfo {author}
  {\bibfnamefont {Paolo}\ \bibnamefont {Umari}}, \ and\ \bibinfo {author}
  {\bibfnamefont {Renata~M.}\ \bibnamefont {Wentzcovitch}},\ }\bibfield
  {title} {\enquote {\bibinfo {title} {{QUANTUM ESPRESSO}: a modular and
  open-source software project for quantum simulations of materials},}\
  }\href@noop {} {\bibfield  {journal} {\bibinfo  {journal} {J.\ Phys.\
  Condens.\ Matter}\ }\textbf {\bibinfo {volume} {21}},\ \bibinfo {pages}
  {395502} (\bibinfo {year} {2009})}\BibitemShut {NoStop}%
\bibitem [{\citenamefont {Perdew}\ \emph {et~al.}(1996)\citenamefont {Perdew},
  \citenamefont {Burke},\ and\ \citenamefont {Ernzerhof}}]{PBE96}%
  \BibitemOpen
  \bibfield  {author} {\bibinfo {author} {\bibfnamefont {John~P.}\ \bibnamefont
  {Perdew}}, \bibinfo {author} {\bibfnamefont {Kieron}\ \bibnamefont {Burke}},
  \ and\ \bibinfo {author} {\bibfnamefont {Matthias}\ \bibnamefont
  {Ernzerhof}},\ }\bibfield  {title} {\enquote {\bibinfo {title} {Generalized
  gradient approximation made simple},}\ }\href@noop {} {\bibfield  {journal}
  {\bibinfo  {journal} {Phys.\ Rev.\ Lett.}\ }\textbf {\bibinfo {volume}
  {77}},\ \bibinfo {pages} {3865--3868} (\bibinfo {year} {1996})}\BibitemShut
  {NoStop}%
\bibitem [{\citenamefont {Hamann}(2013)}]{H13_3}%
  \BibitemOpen
  \bibfield  {author} {\bibinfo {author} {\bibfnamefont {D.~R.}\ \bibnamefont
  {Hamann}},\ }\bibfield  {title} {\enquote {\bibinfo {title} {Optimized
  norm-conserving vanderbilt pseudopotentials},}\ }\href@noop {} {\bibfield
  {journal} {\bibinfo  {journal} {Phys.\ Rev.\ B}\ }\textbf {\bibinfo {volume}
  {88}},\ \bibinfo {pages} {085117} (\bibinfo {year} {2013})}\BibitemShut
  {NoStop}%
\bibitem [{\citenamefont {Schlipf}\ and\ \citenamefont {Gygi}(2015)}]{SG15}%
  \BibitemOpen
  \bibfield  {author} {\bibinfo {author} {\bibfnamefont {Martin}\ \bibnamefont
  {Schlipf}}\ and\ \bibinfo {author} {\bibfnamefont {Fran{\c{c}}ois}\
  \bibnamefont {Gygi}},\ }\bibfield  {title} {\enquote {\bibinfo {title}
  {Optimization algorithm for the generation of {ONCV} pseudopotentials},}\
  }\href@noop {} {\bibfield  {journal} {\bibinfo  {journal} {Comput.\ Phys.\
  Commun.}\ }\textbf {\bibinfo {volume} {196}},\ \bibinfo {pages} {36--44}
  (\bibinfo {year} {2015})}\BibitemShut {NoStop}%
\bibitem [{\citenamefont {Marini}\ \emph {et~al.}(2009)\citenamefont {Marini},
  \citenamefont {Hogan}, \citenamefont {Gr{\"u}ning},\ and\ \citenamefont
  {Varsano}}]{MHG09}%
  \BibitemOpen
  \bibfield  {author} {\bibinfo {author} {\bibfnamefont {A.}~\bibnamefont
  {Marini}}, \bibinfo {author} {\bibfnamefont {C.}~\bibnamefont {Hogan}},
  \bibinfo {author} {\bibfnamefont {M.}~\bibnamefont {Gr{\"u}ning}}, \ and\
  \bibinfo {author} {\bibfnamefont {D.}~\bibnamefont {Varsano}},\ }\bibfield
  {title} {\enquote {\bibinfo {title} {Yambo: an ab initio tool for excited
  state calculations},}\ }\href@noop {} {\bibfield  {journal} {\bibinfo
  {journal} {Comput.\ Phys.\ Commun.}\ }\textbf {\bibinfo {volume} {180}},\
  \bibinfo {pages} {1392--1403} (\bibinfo {year} {2009})}\BibitemShut {NoStop}%
\bibitem [{\citenamefont {Krauss}\ \emph {et~al.}(2018)\citenamefont {Krauss},
  \citenamefont {Kullock}, \citenamefont {Wu}, \citenamefont {Geisler},
  \citenamefont {Lundt}, \citenamefont {Kamp},\ and\ \citenamefont
  {Hecht}}]{KKW18}%
  \BibitemOpen
  \bibfield  {author} {\bibinfo {author} {\bibfnamefont {Enno}\ \bibnamefont
  {Krauss}}, \bibinfo {author} {\bibfnamefont {René}\ \bibnamefont {Kullock}},
  \bibinfo {author} {\bibfnamefont {Xiaofei}\ \bibnamefont {Wu}}, \bibinfo
  {author} {\bibfnamefont {Peter}\ \bibnamefont {Geisler}}, \bibinfo {author}
  {\bibfnamefont {Nils}\ \bibnamefont {Lundt}}, \bibinfo {author}
  {\bibfnamefont {Martin}\ \bibnamefont {Kamp}}, \ and\ \bibinfo {author}
  {\bibfnamefont {Bert}\ \bibnamefont {Hecht}},\ }\bibfield  {title} {\enquote
  {\bibinfo {title} {Controlled growth of high-aspect-ratio single-crystalline
  gold platelets},}\ }\href {\doibase 10.1021/acs.cgd.7b00849} {\bibfield
  {journal} {\bibinfo  {journal} {Cryst.\ Growth\ Des.}\ }\textbf {\bibinfo
  {volume} {18}},\ \bibinfo {pages} {1297--1302} (\bibinfo {year}
  {2018})}\BibitemShut {NoStop}%
\bibitem [{\citenamefont {Kiani}\ and\ \citenamefont {Tagliabue}(2022)}]{KT22}%
  \BibitemOpen
  \bibfield  {author} {\bibinfo {author} {\bibfnamefont {Fatemeh}\ \bibnamefont
  {Kiani}}\ and\ \bibinfo {author} {\bibfnamefont {Giulia}\ \bibnamefont
  {Tagliabue}},\ }\bibfield  {title} {\enquote {\bibinfo {title} {High aspect
  ratio {Au} microflakes via gap-assisted synthesis},}\ }\href {\doibase
  10.1021/acs.chemmater.1c03908} {\bibfield  {journal} {\bibinfo  {journal}
  {Chem.\ Mater.}\ }\textbf {\bibinfo {volume} {34}},\ \bibinfo {pages}
  {1278--1288} (\bibinfo {year} {2022})}\BibitemShut {NoStop}%
\end{thebibliography}


\pagebreak

\begin{widetext}

\section{SUPPLEMENTARY FIGURES}

\pagebreak

\begin{figure}
    \centering
    \includegraphics[width=1\textwidth]{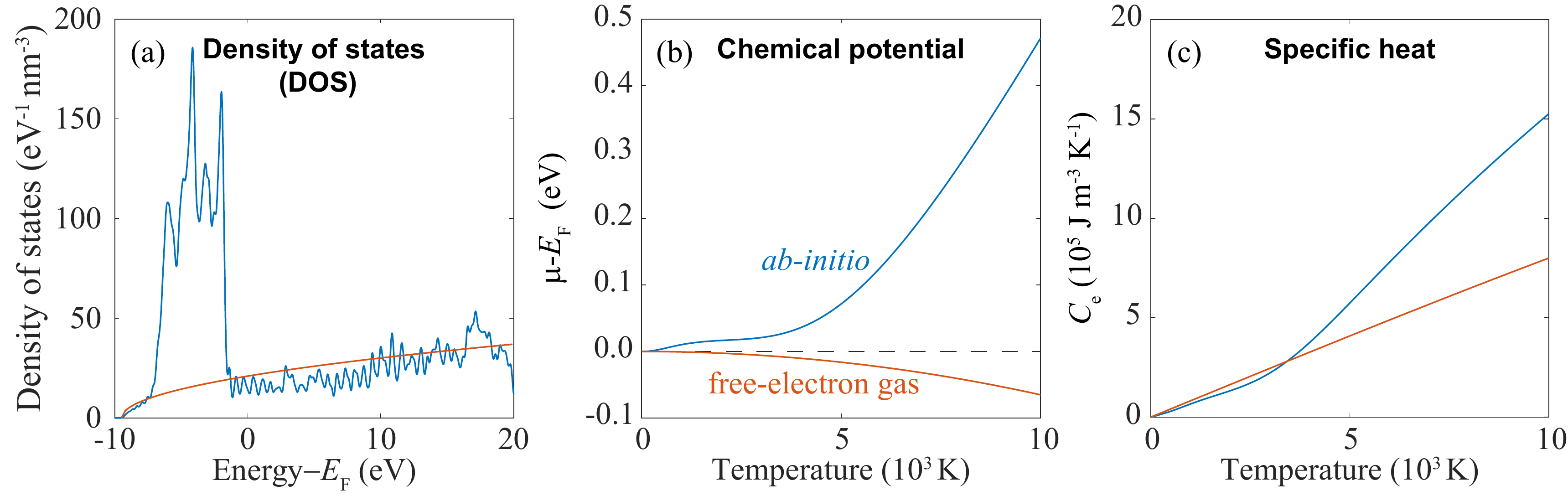}
    \caption{{\bf Electronic and thermal properties of bulk gold}. We compare results obtained from the free-electron model \cite{AM1976} (red curves) and from the electronic band structure calculated by using Quantum Espresso (blue curves). (a) Density of states per unit of volume and electron energy. 
    (b) Temperature dependence of the chemical potential relative to its value at zero temperature (i.e., the Fermi energy $E_{\rm F}$). 
    (c) Specific heat.}
    \label{FigS1}
\end{figure}

\begin{figure}
    \centering
    \includegraphics[width=1\textwidth]{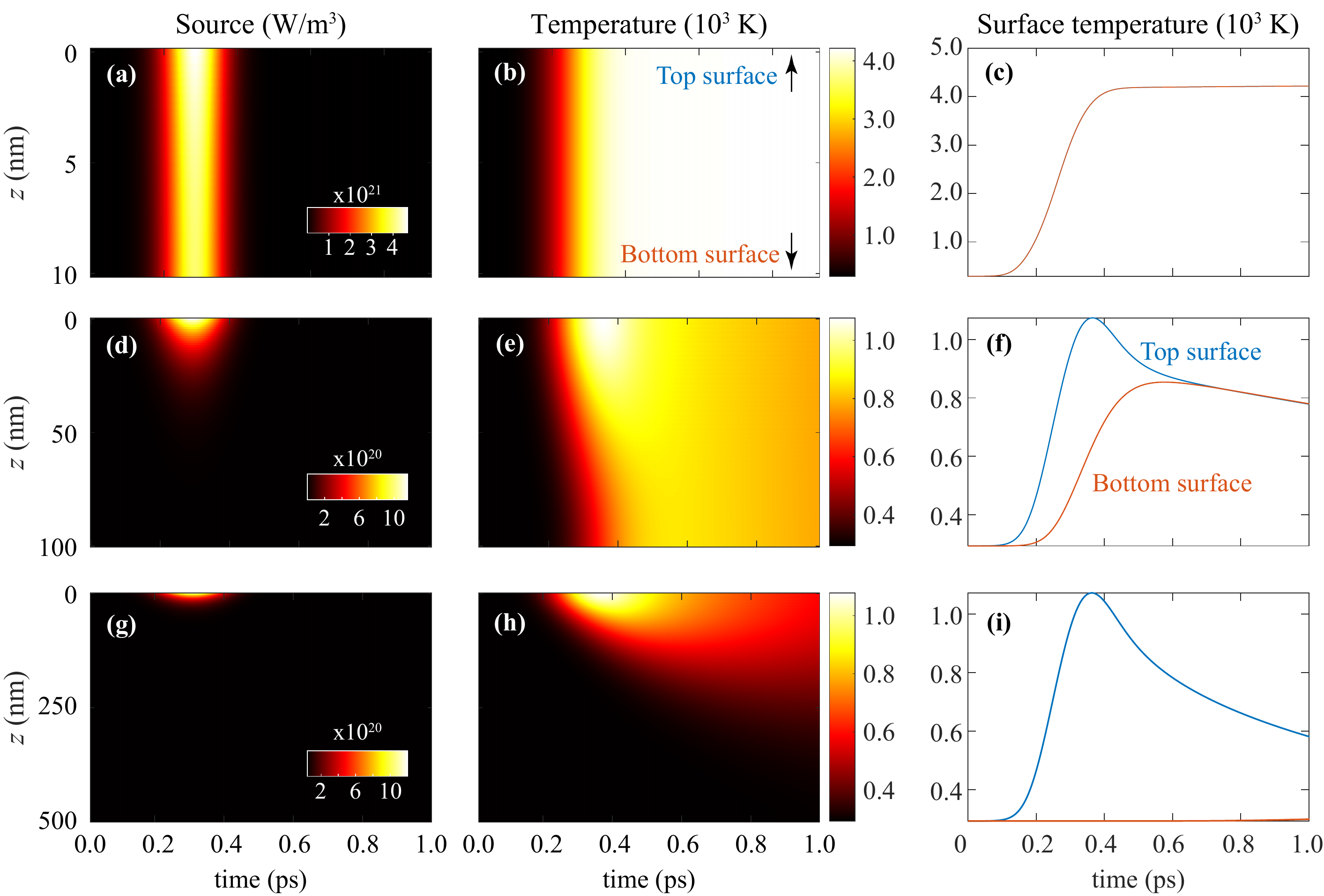}
    \caption{{\bf Evolution of the electronic temperature in optically-pumped gold films}. Heat dynamics for an ultrathin film of 10~nm thickness (a-c), a medium-size film of 100~nm thickness (d-f), and a thicker film (500~nm) resembling a semi-infinite medium (g-i). We consider irradiation by a normally-incident 150~fs Gaussian pulse of 10~mJ/cm$^2$ fluence and 800~nm central wavelength. Panels (a,d,g) show the power density absorbed from the optical pump as a function of time (horizontal axis) and depth (vertical axis). Panels (b,e,h) show the time- and depth-dependence of the resulting electronic temperature. Panels (c,f,i) show the temporal evolution of the temperature at the top (facing the illumination) and bottom film surfaces, as extracted from (b,e,h). The ultrathin film (a-c) has a nearly uniform distribution of power absorption and temperature as a function of depth. Absorption does not reach beyond the skin depth in thicker films, from which thermal diffusion extends the temperature increase towards deeper regions, and eventually relaxation to the lattice causes a decline in temperature.}
    \label{FigS2}
\end{figure}

\begin{figure}
    \centering
    \includegraphics[width=1.0\textwidth]{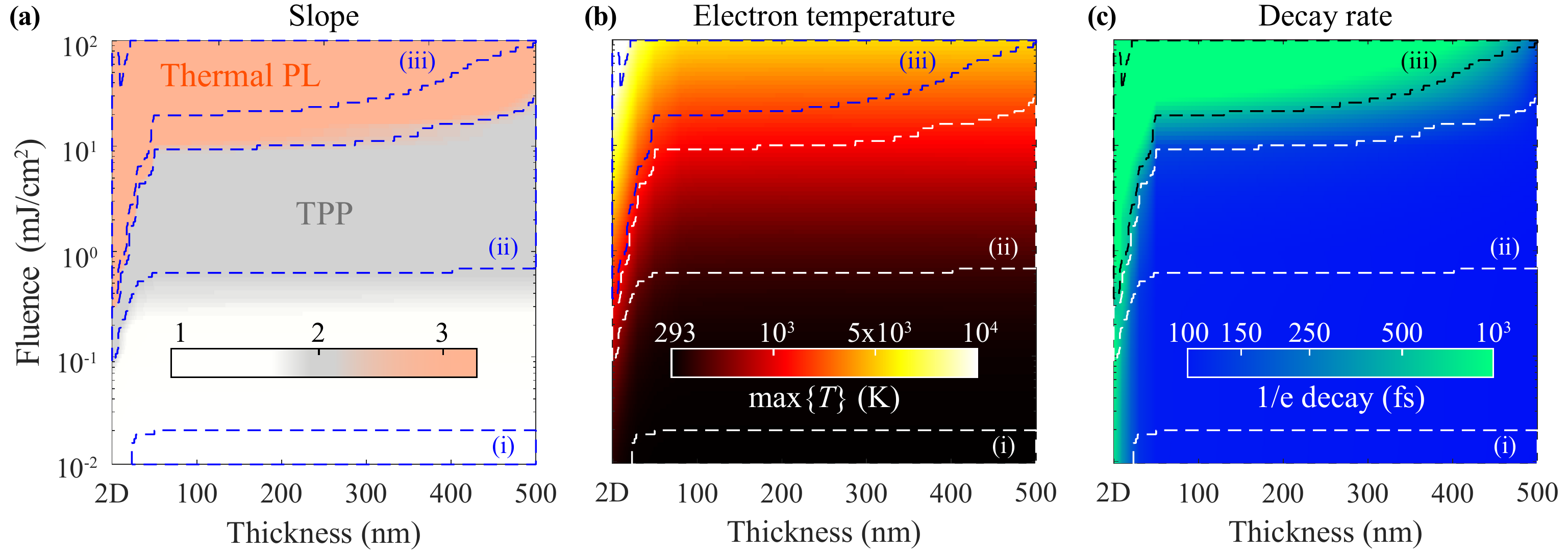}
    \caption{{\bf Photoluminescence dependence on fluence and thickness.} Using information from Fig.~\ref{Fig3} in the main text, panel (a) characterizes the slope of the evolution of PL as a function of thickness and fluence. Shaded areas indicate the NPL and PL regions, where the dashed lines labeled with roman numbers (i), (ii), and (iii) mark the contours for which the slopes are 1, 2, and 3, respectively, coordinated with the corresponding marks in Fig.~\ref{Fig3}. Similarly, panel (b) shows the maximum electron temperature, while (c) depicts the fitting of the decay rate of the normalized intensity in the pump-probe simulations, both of them in logarithmic scale.}
    \label{FigS3}
\end{figure}

\begin{figure}
    \centering
    \includegraphics[width=0.4\textwidth]{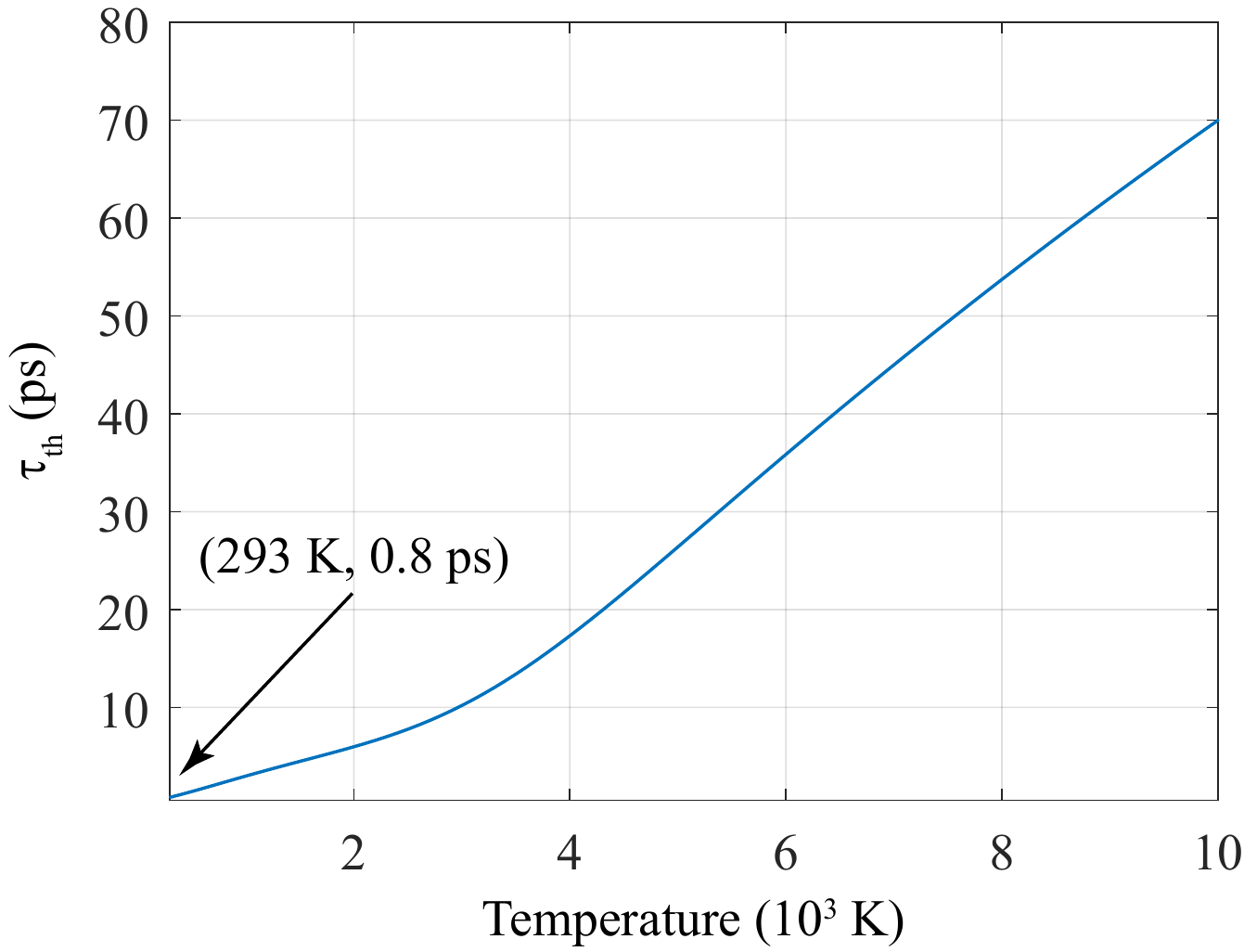}
    \caption{{\bf Thermal characteristic time.} We plot $\tau_{\rm th} = C_e(T)/G$ as a function of electronic temperature, with the heat capacity $C_e(T)$ taken from Fig.~\ref{FigS1} and $G=2.2 \times 10^{16}$\,W\,m$^{-3}$K$^{-1}$ \cite{LKL11}.}
    \label{FigS4}
\end{figure}

\begin{figure}
    \centering
    \includegraphics[width=0.9\textwidth]{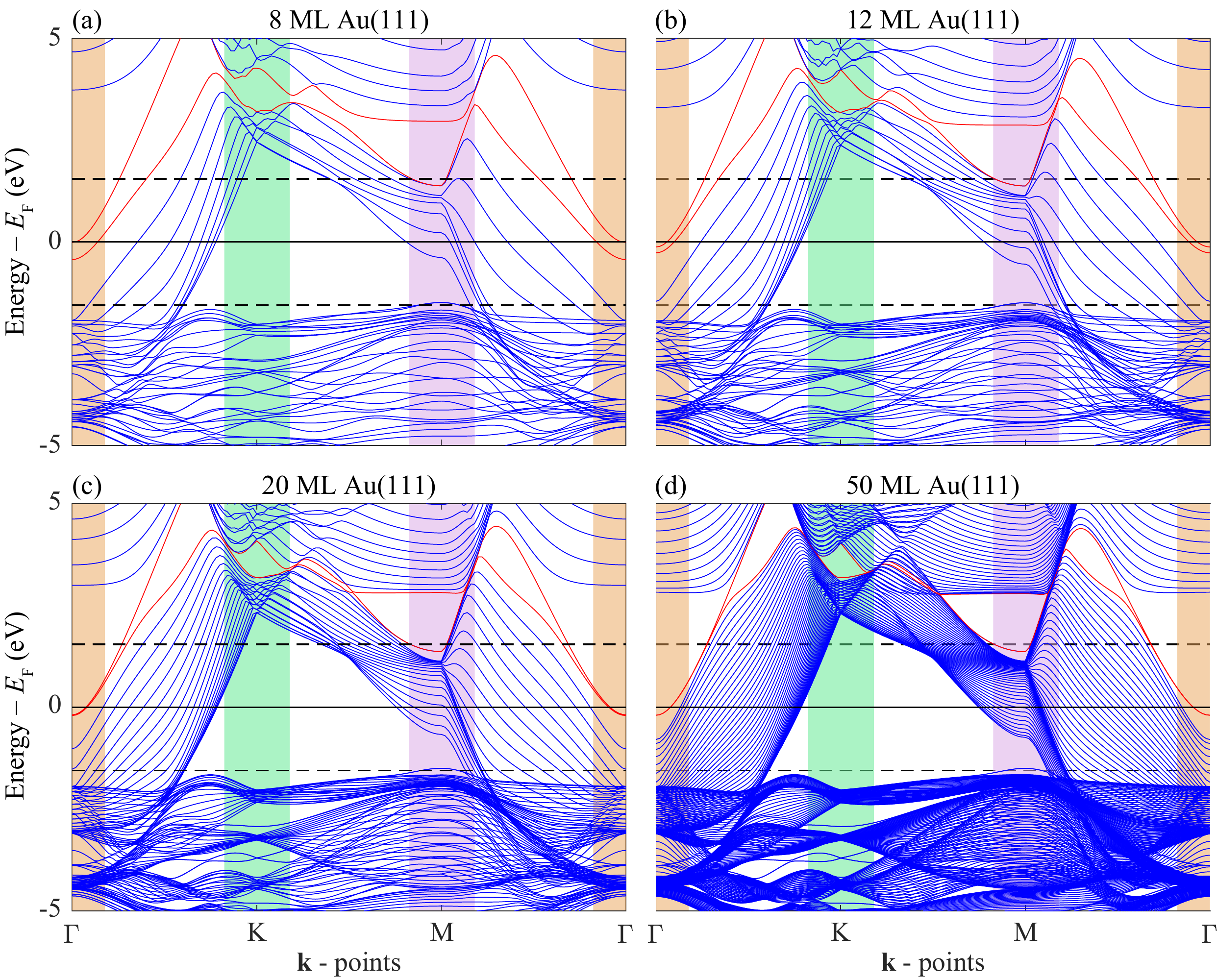}
    \caption{{\bf Band structure of few-monolayer Au(111) films.} Electronic band structure of Au(111) films consisting of (a) 8 ML, (b) 12 ML, (c) 20 ML, and (d) 50 ML. The red lines highlight the surface states. The shaded areas indicates the zones near the $\Gamma$ (beige color), K (green), and M (light purple) points. The dashed black line fixes an energy $\pm \hbar\omega_\inn \approx \pm 1.55$~eV above and below the Fermi energy respectively.}
    \label{FigS5}
\end{figure}

\begin{figure}
    \centering
    \includegraphics[width=.9\textwidth]{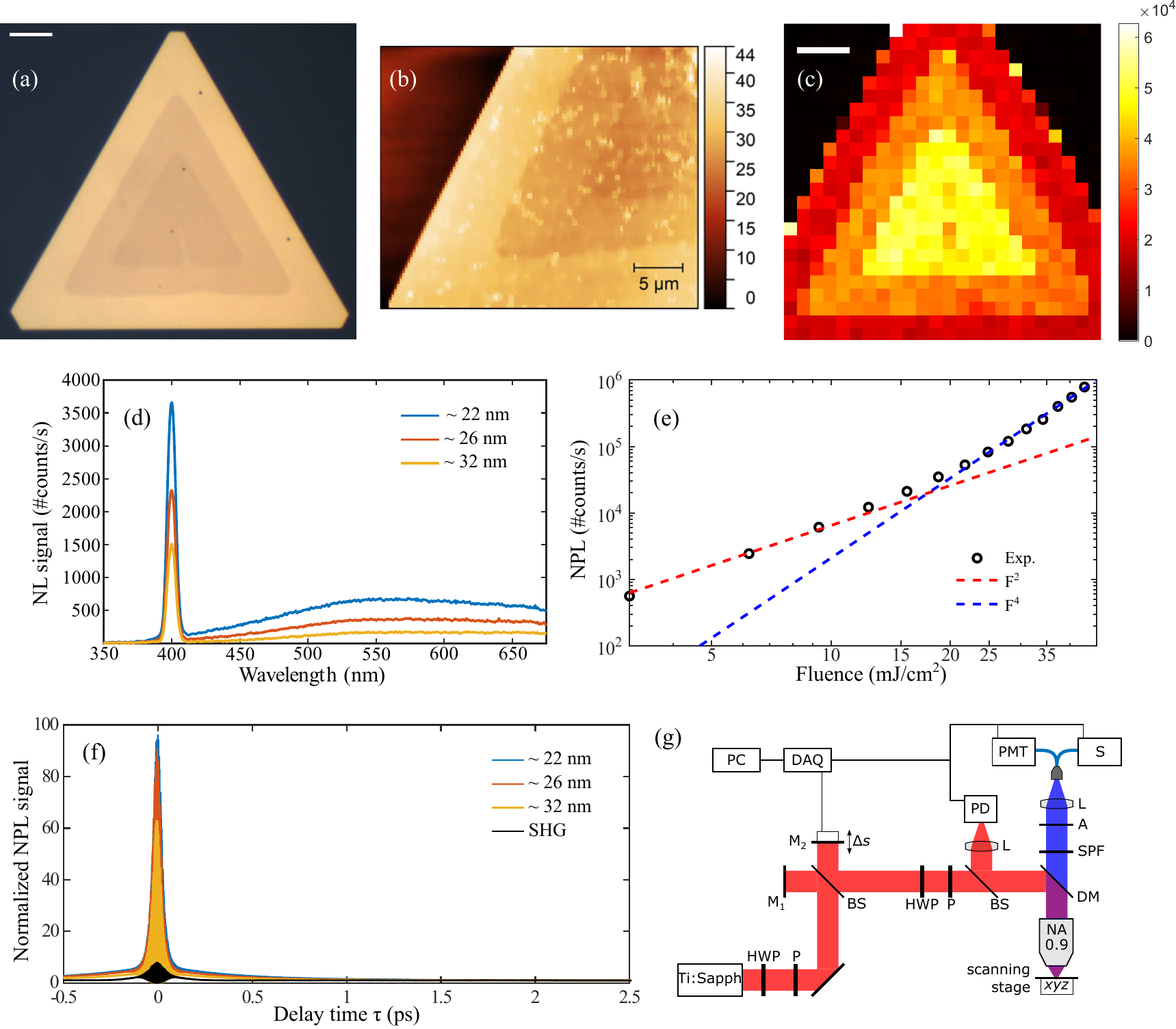}
    \caption{{\bf Spectroscopy and time-resolved NPL measurements of a gold flake} (a) White-light reflection optical micrograph (scale bar: \SI{10}{\micro\meter}). (b) Atomic force microscope (AFM) topography map of the sample region. (c) NPL confocal scan image of the sample (scale bar: \SI{10}{\micro\meter}). (d) Nonlinear spectra acquired at three positions corresponding to three different thicknesses of the gold flake. (e) Power-dependence measurements acquired at the center of the flake (thickness of $\approx\SI{22}{\nano\meter}$). (f) Full-range (i.e. zoom-out of the curves shown in Fig.~\ref{Fig5} in the main text) plot of the two-pulse correlation measurements at the same three positions as in (d) compared with a two-pulse correlation trace acquired from a LiNBO$_3$ crystal (SHG). (g) Experimental setup used to perform spectroscopy and two-pulse correlation measurements, consisting of the following components: excitation laser source (Ti:Saph mode-locked oscillator, approx.~\SI{120}{\pico\second} pulsed output, Tsunami 3941 by Spectra-Physics); half-wave plate (HWP); polarizer (P); beam splitter (BS); fixed mirror (M$_1$); moving mirror mounted on a motorized translation stage (M$_2$); dichroic mirror (DM); band-pass filter (BPF); lens (L); single-photon detector (photomultiplying tube, PMT); spectrometer (S); reflected light detector (silicon photodiode, PD).
    \label{FigS6}}
\end{figure}

\end{widetext}

\end{document}